\documentclass[aip,cha,superscriptaddress,showpacs,floatfix,twocolumn,amsmath,amssymb]{revtex4}
\usepackage[]{graphicx} 
\usepackage{amsmath}
\usepackage{amssymb}
\usepackage{epsfig}
\usepackage{psfrag}

\def\be{\begin{equation}}
\def\ee{\end{equation}}
\def\bea{\begin{eqnarray}}
\def\eea{\end{eqnarray}}

\newcommand{\refsec}[1]{\mbox{Sec.~\ref{#1}}}

\newcommand{\cT}{$\mathcal{T}$}

\begin{document}

\hyphenation{re-so-nan-ce re-so-nan-ces ex-ci-ta-tion z-ex-ci-ta-tion di-elec-tric ap-pro-xi-ma-tion ra-dia-tion Me-cha-nics quan-tum pro-posed Con-cepts pro-duct Reh-feld ob-ser-va-ble Se-ve-ral rea-so-nable Ap-pa-rent-ly re-pe-ti-tions re-la-tive quan-tum su-per-con-duc-ting ap-pro-xi-mate cri-ti-cal func-tion wave-guide wave-guides}

\title{Quantum and Wave Dynamical Chaos in Superconducting Microwave Billiards}
\author{B. Dietz}
\email{dietz@ikp.tu-darmstadt.de}
\affiliation{Institut f\"ur Kernphysik, Technische Universit\"at Darmstadt, D-64289 Darmstadt, Germany}
\author{A. Richter}
\email{richter@ikp.tu-darmstadt.de}
\affiliation{Institut f\"ur Kernphysik, Technische Universit\"at Darmstadt, D-64289 Darmstadt, Germany}

\begin{abstract}
Experiments with superconducting microwave cavities have been performed in our laboratory for more than two decades. The purpose of the present article is to recapitulate some of the highlights achieved. We briefly review (i) results obtained with flat, cylindrical microwave resonators, so-called microwave billiards, concerning the universal fluctuation properties of the eigenvalues of classically chaotic systems with no, a threefold and a broken symmetry; (ii) summarize our findings concerning the wave-dynamical chaos in three-dimensional microwave cavities; (iii) present a new approach for the understanding of the phenomenon of dynamical tunneling which was developed on the basis of experiments that were performed recently with unprecedented precision, and finally, (iv) give an insight into an ongoing project, where we investigate universal properties of (artificial) graphene with superconducting microwave photonic crystals that are enclosed in a microwave resonator, i.e., so-called Dirac billiards. 

\end{abstract}
\maketitle

{\bf Quantum chaos, a field that focuses on the quantum manifestations of classical chaos, attracted a lot of attention during the last $35$ years. The signatures of classical chaos in universal properties of the corresponding quantum system are indeed of relevance in nuclear physics, atomic physics, condensed matter physics, optics and in acoustics. They were observed in the statistical properties of the eigenvalues and the wave functions, in the fluctuation properties of the scattering matrix elements of chaotic scattering processes, in the transport properties of quantum dots and also in systems, where time-reversal invariance is broken, e.g., by a magnetic field. Billiards provide a most appropriate system for the study of features from the field of quantum chaos, because the degree of chaoticity of their classical dynamics only depends on their shape. The eigenvalues and wave functions of two-dimensional quantum billiards are measured since two decades in our laboratory using flat cylindrical, superconducting and normal conducting microwave resonators, respectively. Thereby large sequences of approximately 1000 eigenvalues were determined in the first experiments with a hitherto never achieved high precision yielding new insight into quantum chaos phenomena, because at that time their computation was either not possible or only feasible with a large numerical effort. We will present the results of these and subsequent experiments and also concerning our latest studies of dynamical tunneling. The latter demonstrate that even nowadays the measurements with superconducting microwave resonators may provide valuable new information that is indispensable for the development of a theoretical description of quantum phenomena like dynamical tunneling, even though eigenvalues may now be computed with high accuracy. Most of the works on quantum chaos concentrate on quantum systems described by the non-relativistic Schr\"odinger equation. Only a few studies focus on the properties of the eigenvalues and the wave functions of relativistic billiards that are governed by the Dirac equation of a massless spin-$1/2$ particle. We started lately with their experimental investigation and will present the system and recent results on the statistical properties of the associated eigenvalues in this article.}

\section{\label{Intro}Introduction}
\begin{figure}[h!]
\includegraphics[width=\linewidth]{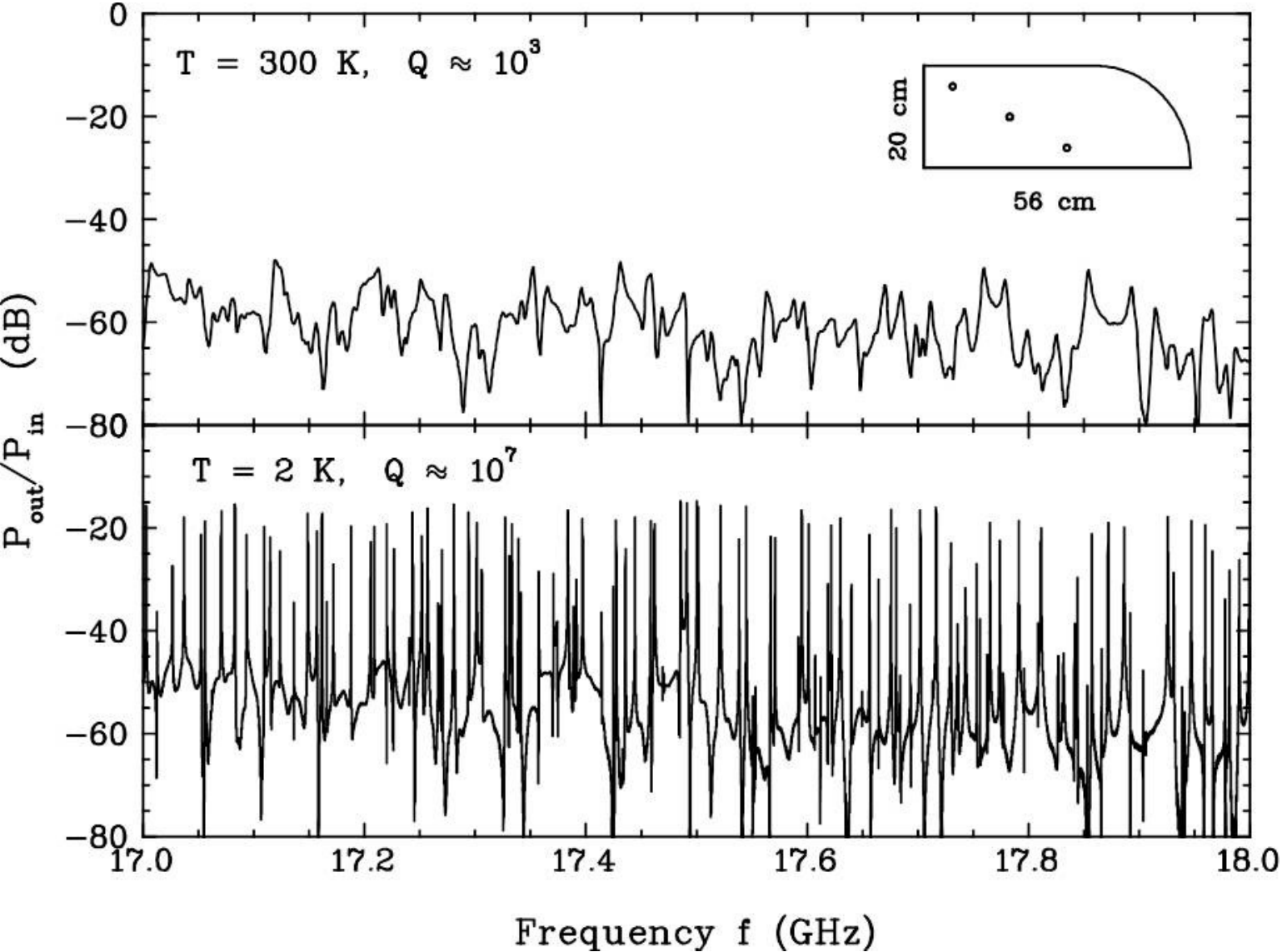}
\caption{Part of the transmission spectrum, i.e., the outcoupled over incoupled power $\rm P_{out}/P_{in}$ versus the excitation frequency, of a microwave billiard with the shape of a quarter stadium measured at room temperature (upper panel) and at 2 K (lower panel), where it is normal conducting and superconducting, respectively. The shape of the resonator and the positions of the antennas are illustrated in the inset. Reprinted from Phys. Rev. Lett. {\bf 69}, 1296 (1992).}
\label{spektren2dstadium}
\end{figure}
This article provides a brief survey on groundbreaking experiments that were performed in the last two decades in our group with superconducting microwave billiards. They focused on the understanding of the generic properties of quantum systems like nuclei and of condensed matter structures like graphene or fullerene. In the first case, we exploited the equivalence of the scalar Helmholtz equation governing flat, metallic microwave resonators, so-called microwave billiards, and the non-relativistic Schr\"odinger equation of the corresponding quantum billiard, to address problems from the fields of quantum chaos and compound nucleus reaction theory~\cite{Mitchell2010}. In the second case, the analogy between the scalar Helmholtz equation describing microwave photonic crystals squeezed between two metal plates or enclosed in a microwave billiard, so-called Dirac billiards, and the Dirac equation was used to experimentally investigate relativistic phenomena occurring in graphene. Some details on the experimental setups for the measurements of resonance spectra at superconducting conditions will be provided in~\refsec{experiments}.  

A classical billiard~\cite{Berry1981} is a bounded domain, in which a pointlike particle moves freely and is reflected elastically on impact with the boundary. Since its dynamics only depends on its shape, such systems are widely used to investigate the effects of the classical dynamics on properties of the associated quantum system~\cite{StoeckmannBuch2000,Bohigas1984,Richter1999,Haake2001}. The correspondence between the latter, i.e., a quantum billiard and a microwave billiard of the same shape, holds below a certain excitation frequency $f$ of the microwaves that are sent into the resonator, $f\leq f_{max}=c/2d$, where the electric field vector is parallel to the cylinder axis. Here, $c$ is the velocity of light and $d$ is the height of the resonator. Originally, this analogy was used to study universal fluctuation properties of the eigenvalues of a quantum billiard with microwave billiards at room temperature~\cite{Stoeckmann1990,Sridhar1991,Stein1992,So1995}. The resolution of the resonances in the spectra, however, was very limited due to the relatively low quality factor $Q\simeq 10^3$ for normal conducting microwave billiards.  
We obtained the eigenvalues of (non) relativistic quantum billiards by measuring resonance spectra to a very high accuracy with superconducting microwave billiards with quality factors of up to $10^7$. This is an indispensable requirement for the determination of complete sequences of several hundreds and even thousands of eigenvalues, as illustrated in Fig.~\ref{spektren2dstadium}. Shown are transmission spectra of a microwave billiard made out of niobium metal with the shape of a quarter stadium (see~\refsec{2dstadium}) measured at room temperature (upper panel), where the resonator is normal conducting and at $2$~K, where it is superconducting. Its eigenfrequencies, and thus the eigenvalues of the corresponding quantum billiard, are obtained from the positions of the resonances. In the upper spectrum in Fig.~\ref{spektren2dstadium} many resonances overlap, whereas all resonances are well resolved in the lower one, so that the determination of the complete sequence of eigenvalues is achievable only in the superconducting  case. Because of this capability we were able to tackle a wide range of problems that are far from feasible with normal conducting microwave billiards. 

Since the resonance spectra are measured by emitting microwave power into the resonator via one antenna and receiving it at the same or another one, microwave billiards can also be viewed as scattering systems. We utilized this capability and  studied the universal properties of quantum scattering systems with an unprecedented accuracy. Due to limitation in space we focus here only on a few of our experiments. In~Secs. \ref{2dstadium} and~\ref{triangular} we present results on the spectral properties of classically chaotic systems. In~\refsec{symmetrbr} we review our studies concerning the effect of symmetry breaking on the properties of the eigenvalues and the wave functions of chaotic systems. We also investigated the spectral properties of three-dimensional resonators that do not have a quantum analogue. These experiments provided additional important information to previously performed ones~\cite{Schroeder1987,Weaver1989,Deus1995,Ellegaard1995} of, due to the superconductivity, hitherto never achieved precision. One example will be discussed in~\refsec{3d}.  

The analysis of the spectral properties addresses the manifestation of the gross structure of the Poincar\'e surface of section (PSOS), namely the relative sizes of the regions of regular and chaotic dynamics, in the associated quantum system. 
The phenomenon of quantum dynamical tunneling~\cite{Keshavamurthy2011}, that is, the classically forbidden transition between states localized in separated parts of the classical phase space has been studied theoretically intensively, but only in a few experiments~\cite{Dembowski2000a,Baecker2008,Dietz2014}, because the measurements require an exceptionally high precision. In Sec.~\ref{tunneling} we will review the main results of our investigations of chaos-assisted tunneling~\cite{Dembowski2000a} and of dynamical tunneling~\cite{Dietz2014}. 

Only lately we started experimental investigations of relativistic phenomena associated with graphene and will present some results of our most recent experiments in~\refsec{Dirac}. 

\section{\label{experiments}Some details of the experimental setups} 

As emphasized in \refsec{Intro} and illustrated in Fig.~\ref{spektren2dstadium} for the determination of complete sequences of eigenvalues of a quantum billiard from the resonance spectra of the corresponding microwave billiard, high-precision measurements are indispensable. Therefore, we used microwave billiards that were superconducting at the temperature of liquid helium $T=4.2$~K. The first billiards were constructed in the CERN workshop in Geneva. They were made from niobium and quality factors $Q\approx 10^5-10^7$ were achieved. In the very first experiments a microwave billiard with the shape of a quarter stadium, reported on in~\refsec{2dstadium}, was inserted into one of the cryostats of the superconducting Darmstadt electron linear accelerator S-DALINAC~\cite{Alrutz-Ziemssen1990} and the measurements were performed at $2$~K. A part of a spectrum is shown in the lower panel of Fig.~\ref{spektren2dstadium}. In the subsequent experiments the microwave billiards were cooled down to $4.2$~K with liquid helium in a separate LHe bath cryostat. The niobium cavities were made from one piece, so it was not possible, e.g., to vary their shapes or to introduce a scatterer into them. Therefore, since about the year $2000$ the resonators were constructed from three copper parts, a bottom plate, a middle one with a hole of the shape of the billiard and a lid. Recently, we have constructed them from two brass plates, where a bassin in the shape of the billiard was milled out of the bottom one. To attain superconductivity all parts of the resonator were lead plated and then squeezed on top of each other tigthly with screws along the boundary of the billiard.

For the measurement of the resonance spectra microwave power was emitted into the resonator via one wire antenna and coupled out either via the same antenna or via a second one, thus yielding a reflection and a transmission spectrum, respectively. A vectorial network analyzer determined the relative phase and amplitude of the output and input signal. This provided the scattering matrix elements $S_{ba}$ for the scattering from antenna $a$ to antena $b$, with $a,b=1,2,\cdots$ depending on the number of antennas introduced into the resonator via holes in the top plates. As already mentioned above, the height $d$ of the resonator determines the maximum frequency $f_{max}=c/2d$ below which the analogy to the quantum billiard of corresponding shape holds. The billiard's size, and thus the number of attainable eigenmodes~\cite{Weyl1912}, is bounded by that of the cryostat. Accordingly, in the first measurements with microwave billiards that had a height $d=7$~mm approximately $1000$ eigenvalues could be determined below $f_{max}\simeq 20$~GHz, the most recent measurements yielded several thousand eigenvalues using resonators with a height of $d=3$~mm which corresponds to a maximum frequency $f_{max}\simeq 50$~GHz. In sections III, VI, VIIamd VIII we will present results from all types of resonators.  

\section{\label{2dstadium} Quantum chaos in closed and open systems: the Bunimovich stadium billiard}
A multitude of the experiments addressed the statistical properties of the eigenvalues and the wave functions of billiards of different shapes. They focused on problems related to the field of quantum chaos, as, e.g. the validity of the “Bohigas-Giannoni-Schmit” conjecture~\cite{McDonald1979,Casati1980,Berry1981,Bohigas1984} for chaotic systems. It states that the spectral fluctuation properties of a chaotic system coincide with those of random matrices from the Gaussian orthogonal ensemble (GOE), if time-reversal (\cT) is conserved, and with those of random matrices from the Gaussian unitary ensemble (GUE), if it is violated. If the classical dynamics is regular they follow the Poisson statistics. There also exist random-matrix theory (RMT) predictions for systems with mixed regular-chaotic dynamics, and for chaotic systems with a partially broken symmetry, as, e.g., the isospin mixing observed in nuclear physics (see~\refsec{symmetrbr}), and with partially violated \cT invariance. 

\begin{figure}[h!]
\includegraphics[width=0.7\linewidth]{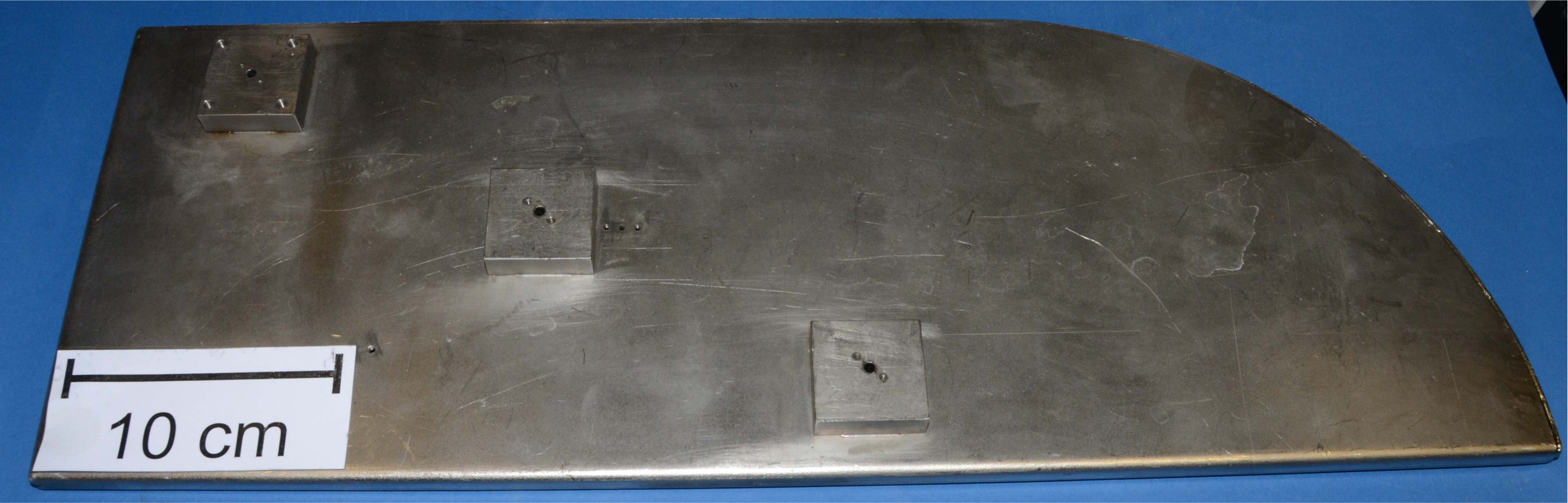}
\caption{(Color online) Photograph of the two-dimensional microwave billiard with the shape of a quarter stadium. The resonator was constructed from niobium which is superconducting below $T_c=9.2$~K. Its height is $8$~mm so the system is governed by the scalar Helmholtz equation below the frequency $f_{max}\approx 18.9$~GHz. The microwave billiard was cooled down to $2$~K in the superconducting Darmstadt electron linear accelerator S-DALINAC and quality factors of $Q\approx 10^5-10^7$ were achieved.}
\label{photo2dstadium}
\end{figure}

The first experiments were performed with a niobium microwave billiard with the shape of a quarter Bunimovich stadium~\cite{Graf1992} shown in Fig.~\ref{photo2dstadium}. The radius of the quarter circle equaled $r=200$~mm and the length of the rectangular part $a=560$~mm. 

\begin{figure}[h!]
\includegraphics[width=0.8\linewidth]{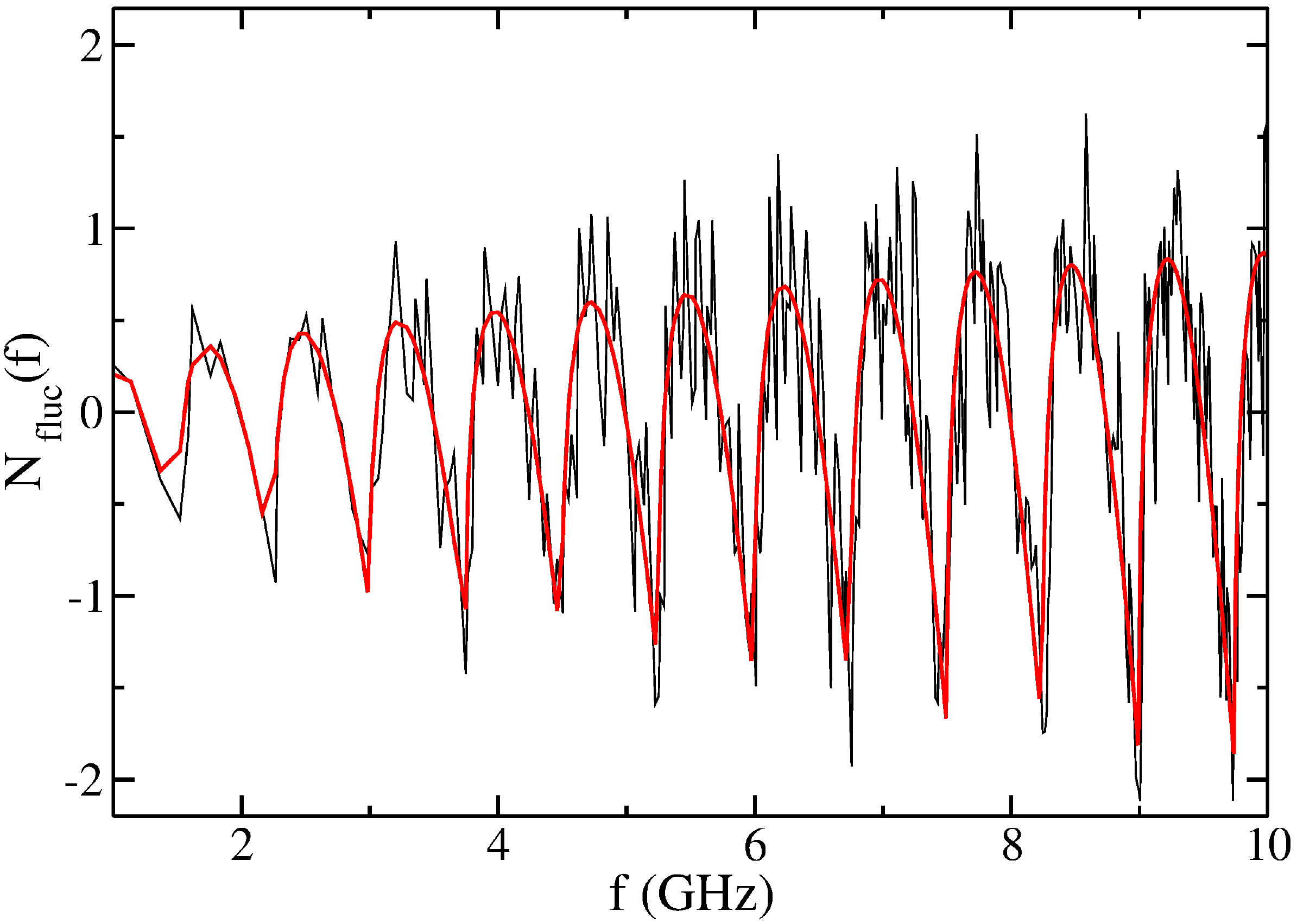}
\caption{(Color online) Fluctuating part of the integrated density of the resonance frequencies obtained for the microwave billiard in Fig.~\ref{photo2dstadium} (black line) and the contributions of the bouncing ball orbits (red line) deduced from Eq.~(\ref{semicl}). Reprinted from Phys. Rev. Lett. {\bf 69}, 1296 (1992).}
\label{nfluc2dstadium}
\end{figure}

\begin{figure}[h!]
\includegraphics[width=0.8\linewidth]{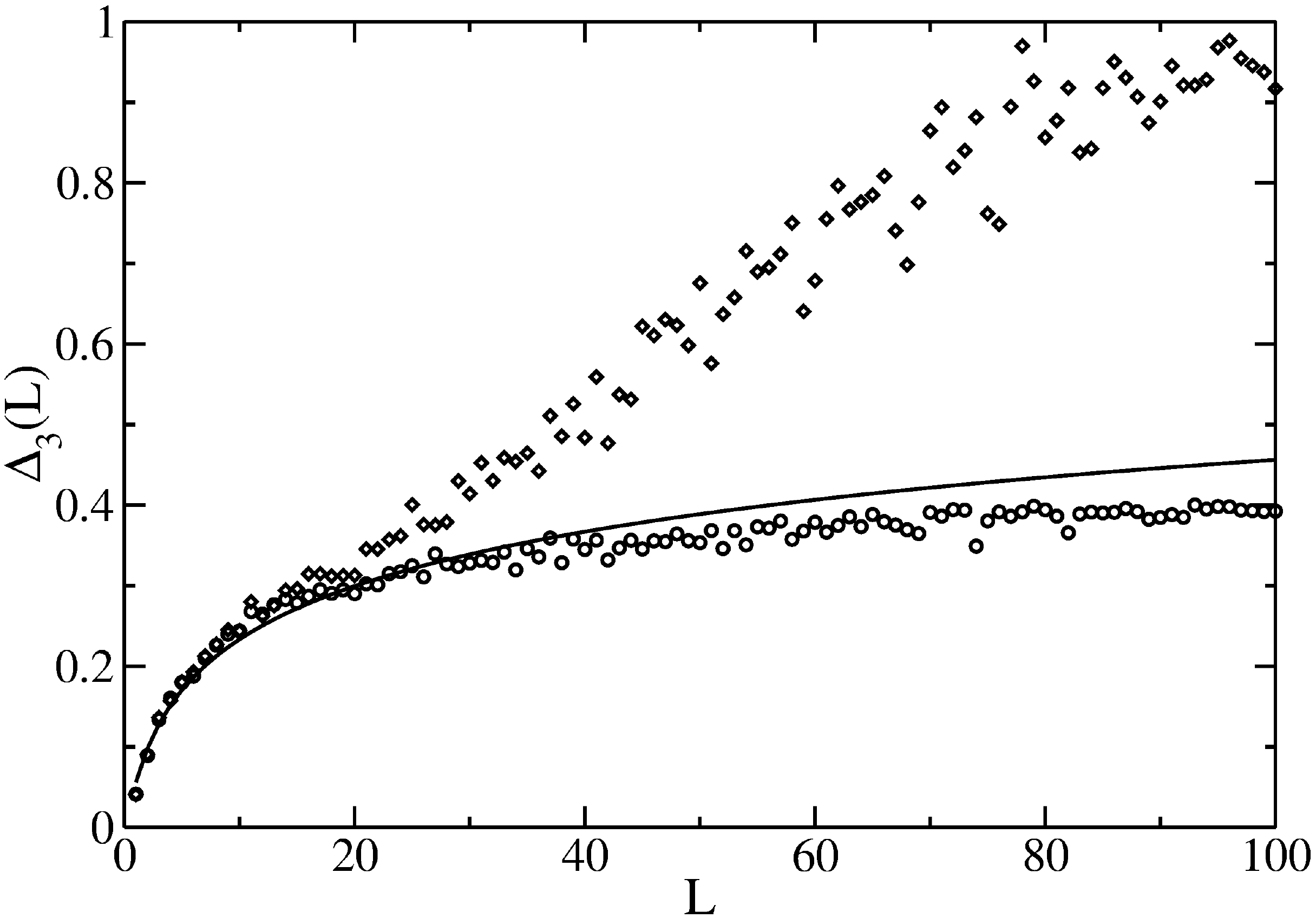}
\caption{The $\Delta_3$ statistics of the resonance frequencies of the microwave billiard in Fig.~\ref{photo2dstadium} before (diamonds) and after (circles) extraction of the contribution from the bouncing-ball orbits compared to the RMT result for the GOE (full line). Reprinted from Phys. Rev. Lett. {\bf 69}, 1296 (1992).}
\label{delta32dstadium}
\end{figure}

A complete sequence of $1060$ resonance frequencies was identified. For the study of the spectral properties the spectra were unfolded to mean resonance spacing unity with Weyl's formula~\cite{Weyl1912} which provides an analytical expression for the smooth part of the integrated resonance density $N(f)$ in terms of a polynomial of second order, $N_{smooth}(f)=\pi/(4c^2)Af^2-B/(4c)f+C$, with $A$ the area and $B$ the perimeter of the billiard. Figure~\ref{nfluc2dstadium} shows the fluctuating part of the integrated resonance density, $N_{fluc}(f)=N(f)-N_{smooth}(f)$ which exhibits fast fluctuations and smooth oscillations. The latter are due to non-generic periodic orbits (POs), so-called bouncing-ball orbits (BBOs), that bounce back and forth between the two parallel side walls of the rectangular part of the stadium billiard. The red (gray) full line shows the analytical result for their contribution,
\begin{equation}
N_{bb}(f)=\frac{a}{2\pi r}\sqrt{2\frac{fr}{c}}\sum_{m=1}^\infty m^{-3/2}\cos\left(4\pi m\frac{fr}{c}-\frac{3\pi}{4}\right).
\label{semicl}
\end{equation}
It was derived on the basis of a semiclassical method~\cite{Sieber1993}. While the bouncing-ball orbits are of measure zero in the classical dynamics they have a strong effect on the spectral propertes of the quantum billiard, as demonstrated in Fig.~\ref{delta32dstadium}. Shown is the Dyson-Mehta $\Delta_3(L)$ statistics~\cite{Mehta1990}, which gives the least-square deviation of the integrated density of the unfolded resonance frequencies from the best straight-line fit in the interval of length $L$. Since the classical dynamics of a Bunimovich stadium is completely chaotic~\cite{Bunimovich1979} it is expected to coincide with the RMT results for the GOE. The circles and the diamonds show the experimental results obtained when unfolding the resonance frequencies $f$ by replacing them as usually by $\tilde f=N_{smooth}(f)$, and by $\tilde f=N_{smooth}(f)+N_{bb}(f)$ taking into account the effect of the BBOs, respectively. It is clearly visible that the agreement between the experimental $\Delta_3$ statistics and the GOE is very good only in the latter case.    

As was noted above, for the measurement of the resonance spectra microwave power is emitted into the resonator via one antenna thereby exciting an electric field mode in its interior and an output signal is received at the same or another antenna. Hence, the resonator is an open scattering system with the antennas acting as single scattering channels. The associated $S$ matrix comprises the scattering between the input and the output antennas and dissipation into the walls, which is accounted for by introducing a large number of weakly coupled fictitious scattering channels. The $S$-matrix formalism for compound nucleus reactions~\cite{Mahaux1969} is equivalent to that for microwave resonators~\cite{Albeverio1996}. We used this analogy to investigate universal properties of the associated $S$ matrix and more generally for quantum scattering processes with intrinsic chaotic dynamics. In regions of isolated resonances, as is the case in the measurements with the superconducting quarter-stadium billiard, the $S$-matrix element for the scattering from antenna $a$ to antenna $b$ is given as
\begin{equation}
S_{ba}(f)=\delta_{ba}-i\frac{\sqrt{\Gamma_{\mu b}\Gamma_{\mu a}}}{f-f_\mu+\frac{i}{2}\Gamma_\mu}\, 
\label{sab}
\end{equation}   
in the vicinity of the resonance $\mu$. Here, $\Gamma_{\mu a}$ and $\Gamma_{\mu b}$ are the partial widths related to the emitting and the receiving antennas $a$ and $b$, respectively, and $\Gamma_\mu$ is the total width of the resonance, $\Gamma_\mu=\sum_{c=1}^3\Gamma_{\mu c}+\Gamma_{\mu ,diss}$. It includes the partial widths of the $3$ antennas attached to the microwave billiard (see Fig.~\ref{photo2dstadium}) and the dissipation in the walls. Note, that only for the spectra that were measured at $2$~K, i.e., for those of the quarter-stadium billiard $\emph{all}$ resonance parameters could be determined from a fit of the Lorentzian in Eq.~(\ref{sab}) to the resonances~\cite{Beck2003}. The reason is, that the contributions $\Gamma_{\mu, diss}$ from the losses due to the dissipation in the walls were negligible as compared to those from the partial widths, whereas this is no longer the case in the measurements at liquid Helium temperatures of $4.2$~K, see below. Figure~\ref{breiten} shows  the total widths $\Gamma_\mu$ and the partial widths $\Gamma_{\mu 2}$ associated with antenna $2$ versus the resonance frequencies $f_\mu$. The data fluctuate strongly around a slow secular variation which is well approximated by a polynomial of $5$th order and was removed for the analysis of their statistical properties by scaling the data with the latter. 

\begin{figure}[h!]
\includegraphics[width=0.8\linewidth]{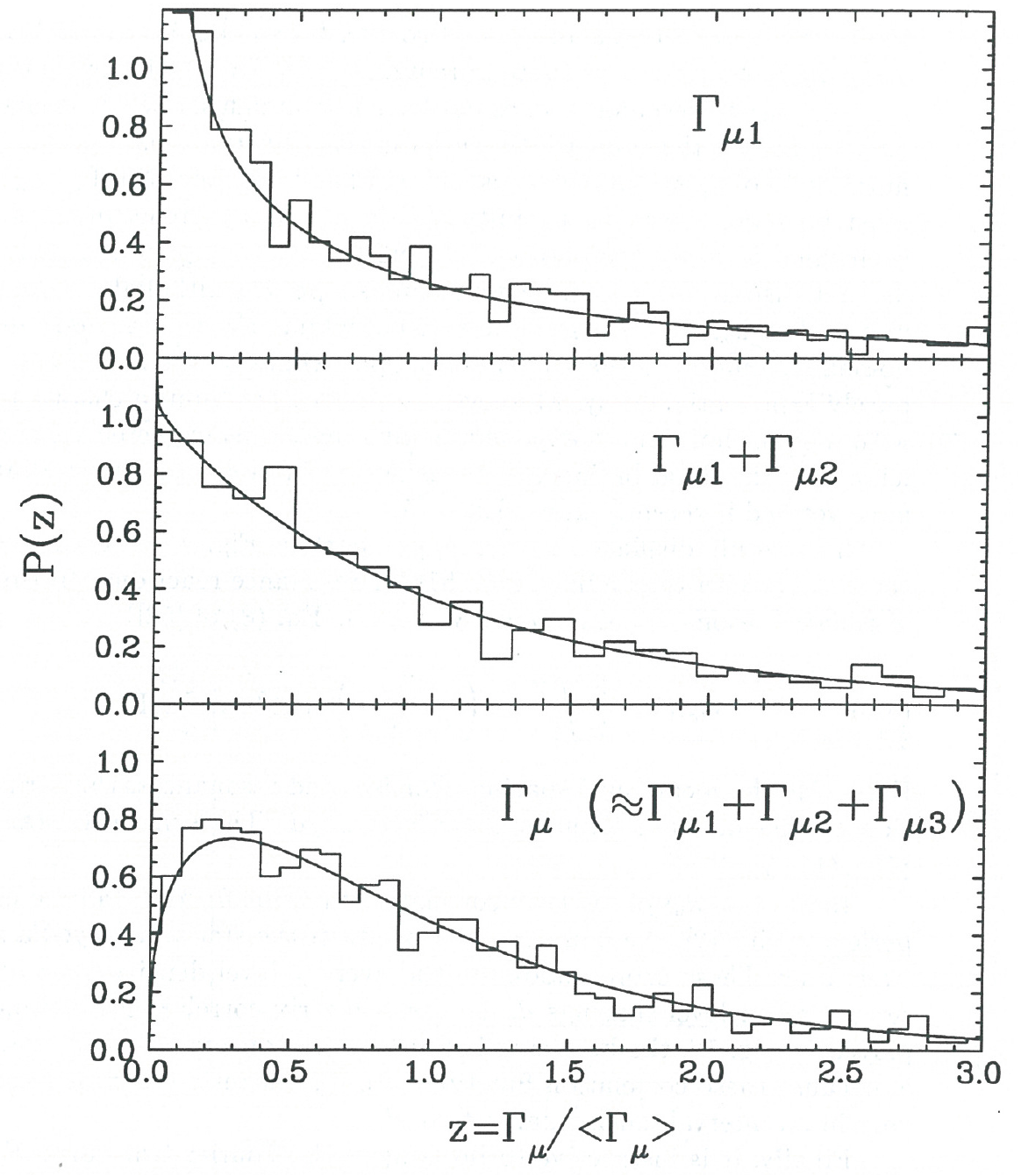}
\caption{Distributions of the partial widths $\Gamma_{\mu a}$ associated with antennas $a=1,2,3$ normalized by their respective mean. Those of $\Gamma_{\mu 1}$ (upper panel) and of the sum of $\Gamma_{\mu 1}+\Gamma_{\mu 2}$ (middle panel) and of all (lower panel) partial widths (histograms) of the resonances measured with the superconducting microwave billiard in Fig.~\ref{photo2dstadium}, respectively, are displayed. The full curves show the $\chi_\nu^2$ distributions of $\nu =1,\, 2$ and $3$ degrees of freedom from top to bottom. Adopted from Ref.~\cite{Richter1999}. With kind permission from Springer Science and Business Media.}
\label{zerfall}
\end{figure}

The distribution of the partial widths is expected to coincide with that of a $\chi_\nu^2$ distribution with the number of degrees of freedom $\nu =1$, which is a Porter-Thomas distribution. The sum of $2$ and $3$ partial widths should coincide with that of $2$ and $3$ degrees of freedom, respectively, if the partial widths indeed are uncorrelated. As demonstrated in Fig.~\ref{zerfall} the agreement between the experimental and the expected results is very good~\cite{Alt1995,Richter1999}.      
\begin{figure}[h!]
\includegraphics[width=0.8\linewidth]{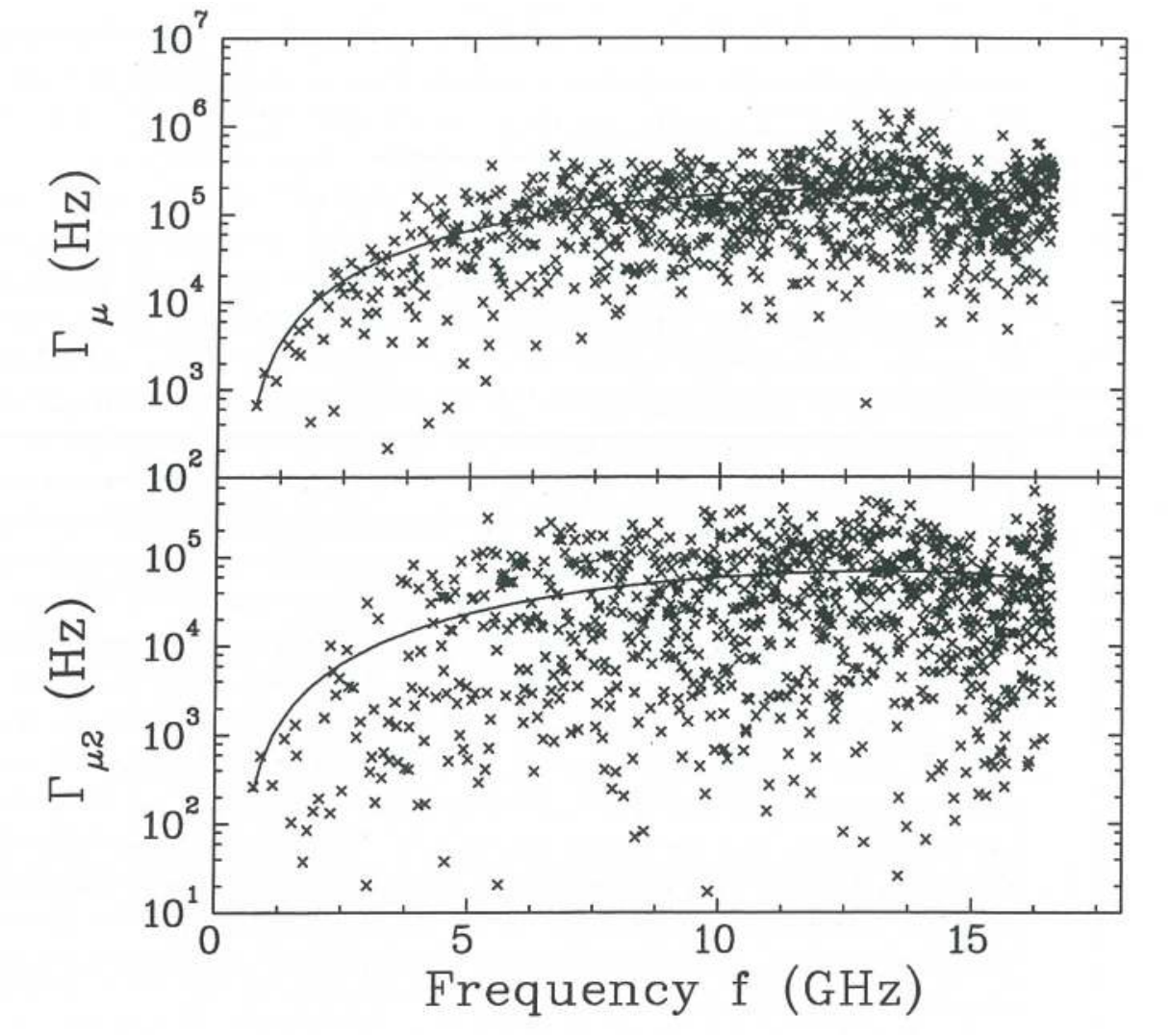}
\caption{The total widths (upper panel) and the partial widths (lower panel) of the resonances measured with the superconducting microwave billiard shown in Fig.~\ref{photo2dstadium} versus their frequencies. The full lines are polynomial fits to the data, that describe the secular dependence of the widths on the frequency. Reprinted from Phys. Rev. Lett. {\bf 74}, 62 (1995).}
\label{breiten}
\end{figure}
Furthermore, the $S$-matrix autocorrelation function was computed using the experimental resonance spectra and the RMT predictions for those of isolated resonances~\cite{Mueller1990} were confirmed. The Fourier tranforms of the $S$-matrix autocorrelation function revealed an algebraic decay. This actually is expected for isolated and weakly overlapping resonances~\cite{Alt1995,Mueller1990}. 

We would like to close this section with a few remarks concerning our experimental tests of analytical results within the theory of chaotic scattering which has been largely developed in the framework of the statistical theory of compound-nucleus reactions~\cite{Mitchell2010}. Predictions for the fluctuation properties of the associated $S$ matrix were obtained based on RMT and on the supersymmetry method~\cite{Verbaarschot1985,Fyodorov2005,Dietz2009a,Kumar2013}. To be more precise, analytic expressions for $S$-matrix correlation functions and the distribution of the $S$-matrix elements have been derived which are valid from the regime of isolated resonances to that of strongly overlapping ones. For their experimental verification, scattering data where taken at room temperature using a microwave resonator with the shape of a chaotic tilted stadium billiard allowing tests also with unprecedented accuracy in the regions of isolated and weakly overlapping resonances~\cite{Dietz2008,Dietz2009a,Dietz2010,Dietz2010a,Dietz2011a}. Experiments were furthermore performed with microwave billiards where a partial \cT violation was induced by a magnetized ferrite~\cite{Dietz2007a}. For the analysis of the latter we extended predictions derived for \cT-invariant systems by Verbaarschot, Weidenm\"uller and Zirnbauer~\cite{Verbaarschot1985} for the $S$-matrix autocorrelation functions to systems with partial \cT violation. Due to the large sets of scattering data measured as function of the excitation frequency we could also study higher order correlation functions, as e.g. the cross-section autocorrelation function to test RMT predictions on cross-section fluctuations~\cite{Dietz2010,Dietz2010a,Dietz2011a}. 
\section{\label{symmetrbr} Study of symmetry breaking with coupled quarter-stadium billiards}
\begin{figure}[h!]
\includegraphics[width=0.8\linewidth]{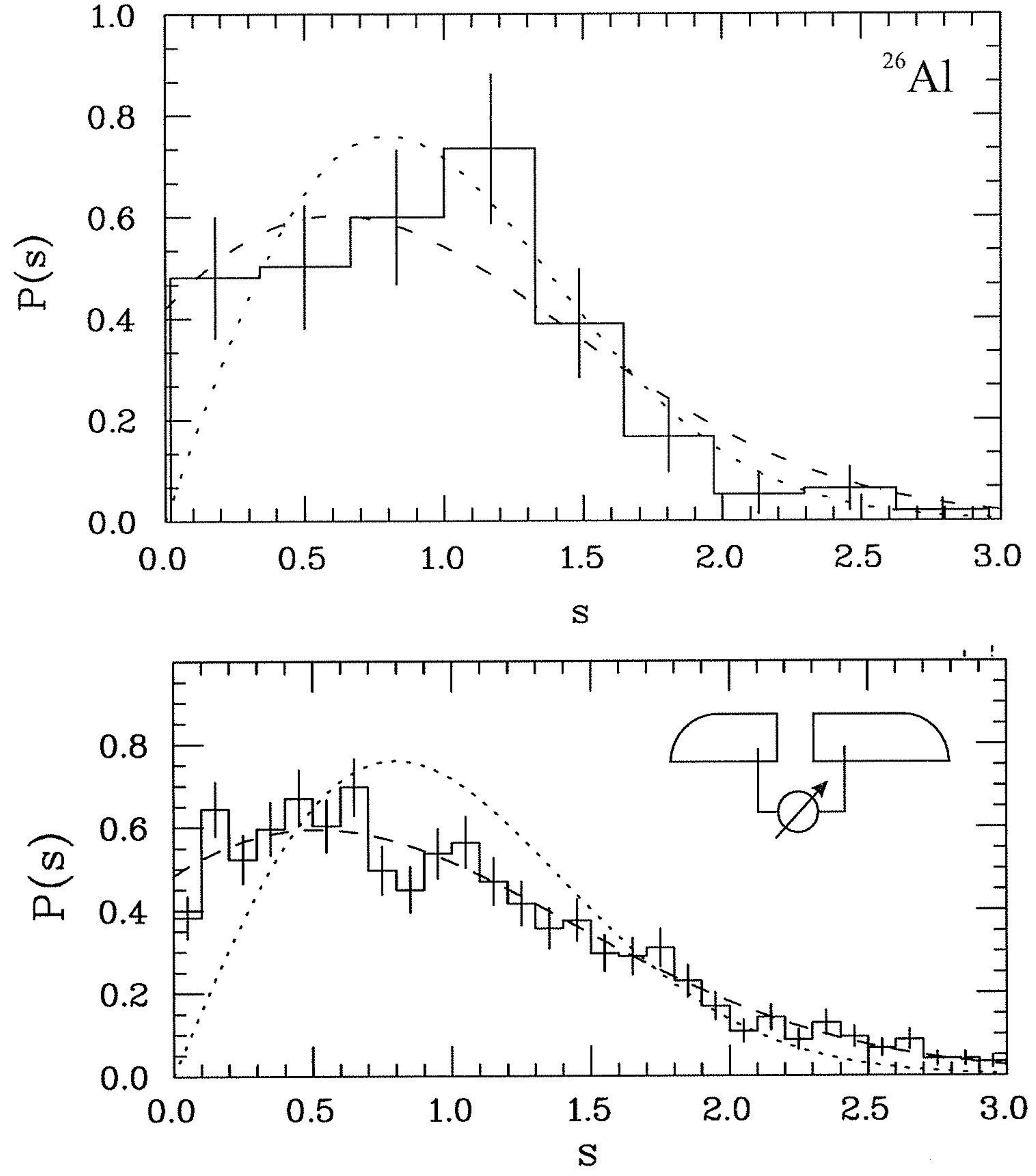}
\caption{Nearest-neighbor spacing distribution of the energies of the first 107 excited states with positive parity of $^{26}$Al (upper panel) and of the resonance frequencies of the uncoupled stadium billiards (lower panel). The dashed lines correpond to the curve for two uncoupled GOEs, the dotted ones to that of one GOE.}
\label{abstgekop}
\end{figure}

Conservation laws, that is, the invariance of physical quantities with respect to certain transformations, play an important role in nearly all fields of physics. In 1918, Emmy Noether formulated a theorem~\cite{Noether1918}, which states that any differentiable symmetry of the action of a physical system has a corresponding conservation law. In a quantum system each conserved quantity corresponds to a quantum number. We investigated the symmetry breaking in a system that is characterized by two quantum numbers, like, e.g., that of the isospin symmetry in $^{26}$Al or $^{30}$P, which is broken due to the Coulomb interaction, see \cite{Mitchell1988,Guhr1990a,Shriner2005} and references cited therein. In $^{26}$Al a \emph{complete} sequence of $107$ states of positive parity from the ground state up to about $8$~MeV excitation energy was identified. Among these $75$ had isospin quantum number $T=0$ and $32$ $T=1$, so one has the situation of two symmetry classes with each characterized by a GOE, since the corresponding classical dynamics is chaotic. The upper panel of Fig.~\ref{abstgekop} shows their nearest-neighbor spacing distribution (histogram). It should coincide with that of two uncoupled GOEs, plotted as dashed line in Fig.~\ref{abstgekop}, if the isospin were a good quantum number. If, on the other hand, the symmetry were completely broken, that is, if there were no isospin symmetry, the distribution should coincide with that of one GOE, shown as dotted line. 
The fact, that the nearest-neighbor spacing distribution coincides with neither of the two clearly indicates that the isospin is partially broken in $^{26}$Al, or equivalently, that the two sets of states with isospin quantum numbers $T=0$ and $T=1$ are mixed. This mixing of the states of two chaotic systems is modelled by an ensemble of random matrices of the form 
\begin{equation}
H^{mixed}=\left({\begin{array}{cc}
                H_1\, &0\\
                0\, &H_2
                \end{array}}
          \right)\, +\lambda D
\left({\begin{array}{cc}
                0\, &V\\
                V^T\, &0
                \end{array}}
          \right)\, ,
\label{eq:mixedH}
\end{equation}
which is a special case of the Rosenzweig-Porter model~\cite{Rosenzweig1960}. Here, the first part preserves the symmetry, where, in the case of $^{26}$Al, $H_1$ and $H_2$ are GOE matrices and describe the spectral properties of the states with $T=0$ and $T=1$, respectively. The symmetry breaking effect of the Coulomb interaction, that is, the mixing of the isospin states is accounted for by the second term, with $D$ denoting the mean level spacing. For a non-vanishing symmetry-breaking parameter $\lambda$ it yields a coupling of $H_1$ and $H_2$. 

\begin{figure}[h!]
\includegraphics[width=0.8\linewidth]{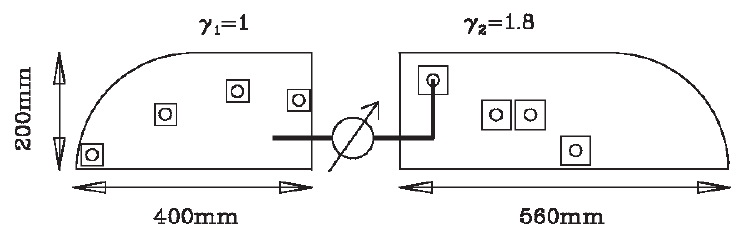}
\caption{Schematic view of the two coupled Bunimovich stadium billiards. The locations of the antennas are marked by circles. Reprinted from Phys. Rev. Lett. {\bf 81}, 4847 (1998).}
\label{gekop}
\end{figure}

The symmetry breaking actually could be tested experimentally by mapping this RMT model onto a system consisting of two coupled Bunimovich stadium billiards~\cite{Alt1998} shown schematically in Fig.~\ref{gekop}.           
A smaller microwave stadium billiard with the ratio $\gamma_1=1.0$ between the length of the rectangle and the radius of the quarter circle was coupled to the one presented in the previous section III and shown in Fig.~\ref{photo2dstadium} which had a ratio $\gamma_2=1.8$. Thus, the number of resonances identified up to $17.5$~GHz in the former, $N_1=608$, is smaller than that in the latter, $N_2=883$. The dynamics in the two uncoupled cavities is simulated in the RMT model [Eq.~(\ref{eq:mixedH})] by GOE matrices $H_1$ and $H_2$ with the dimensions equal to $N_1$ and $N_2$, respectively. For the coupling, the microwave billiards were put on top of each other and a niobium pin, $2$~mm in diameter, penetrated through holes in their walls into both of them. Its size was controlled by the associated penetration length. This led to the formation of a TEM mode between the walls and the pin. The mixing of the $N_1$ and $N_2$ modes of both microwave billiards via the TEM mode was modelled by an $N_1$ dimensional vector $\vec v$ and an $N_2$ dimensional vector $\vec w$ with Gaussian-distributed random numbers as entries, respectively, thus yielding a dyadic structure for $V=\vec v\vec w^T$ in Eq.~(\ref{eq:mixedH}). The distribution of its matrix elements is given by a modified $K_0$ Bessel function. For a more detailed description see Refs.~\cite{Alt1998,Dietz2006a}. 

\begin{figure}[h!]
\includegraphics[width=0.8\linewidth]{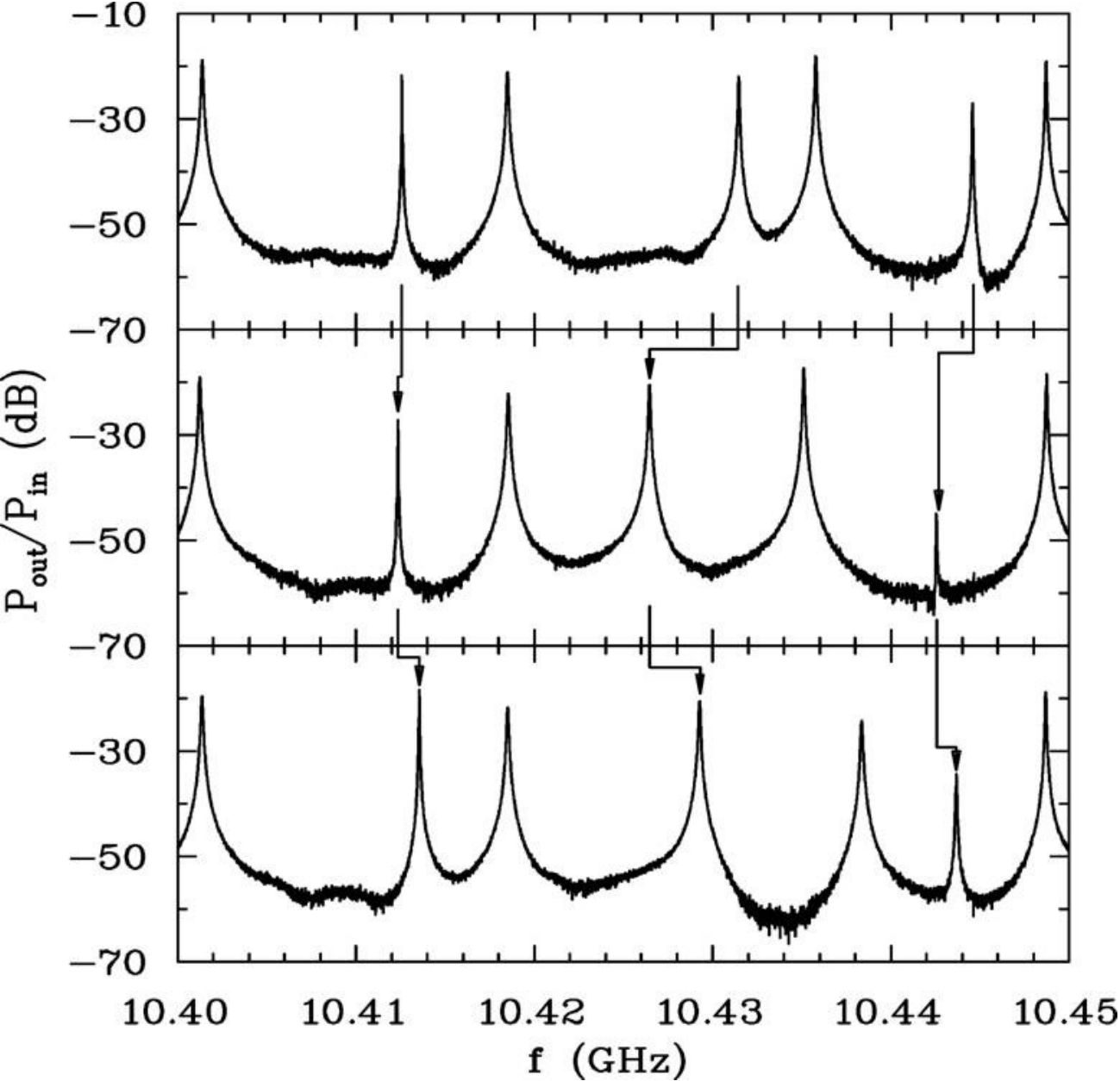}
\caption{Part of the transmission spectra measured with no coupling, with weak coupling and with the strongest achieved coupling from top to bottom. The arrows mark a few resonances that are shifted due to the coupling. Reprinted from Phys. Rev. Lett. {\bf 81}, 4847 (1998).}
\label{spektrengekop}
\end{figure}

Altogether $6$ transmission spectra were measured with three antennas for $5$ different couplings. The effect of the coupling on the spectra is illustrated in Fig.~\ref{spektrengekop}. The change of the coupling led to a shift of resonances, whose associated electric field strength distribution, i.e., wave function, was non-vanishing at the position of the pin. We investigated the influence of the symmetry breaking on the eigenvalues of a chaotic system in terms of the spectral fluctuation properties of the resonance frequencies, and that on the wave functions in terms of the resonance strengths $\Gamma_{\mu a}\Gamma_{\mu b}$ in Eq.~(\ref{sab}). The latter provide information on the electric field strengths, i.e., the wave function components at the positions of the antennas. The lower panel of Fig.~\ref{abstgekop} shows the nearest-neighbor spacing distribution of the unfolded resonance frequencies for the case of no coupling. It is well described by the RMT model [Eq.~(\ref{eq:mixedH})] with the coupling parameter set to zero, $\lambda =0$ (dashed line). 

\begin{figure}[h!]
\includegraphics[width=\linewidth]{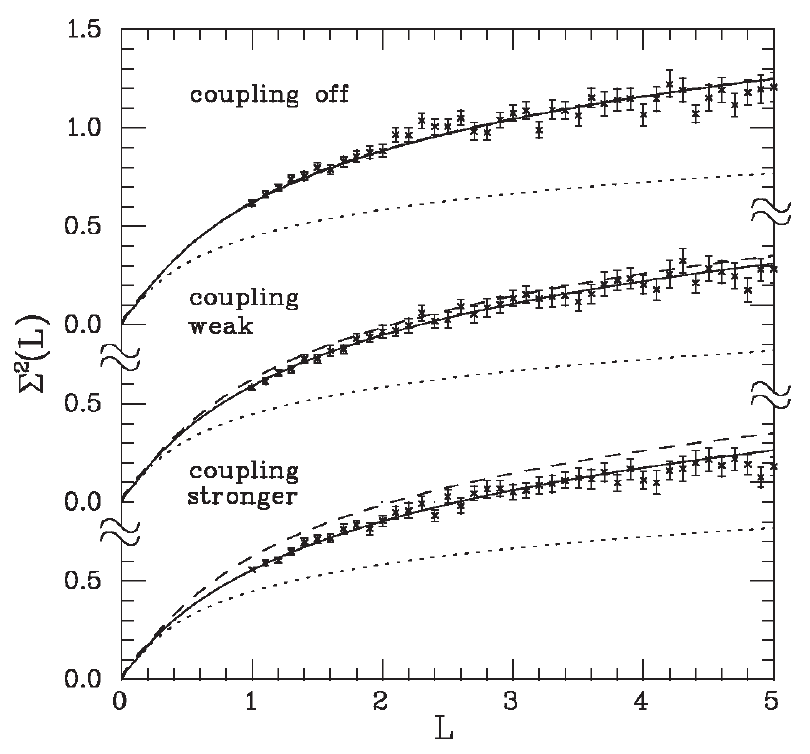}
\caption{The $\Sigma^2$ statistics of the measured resonance frequencies of the coupled stadium billiards for no coupling, a weak coupling and the strongest achieved coupling from top to bottom (filled circles). The dashed and the dotted lines show the RMT results for two uncoupled GOEs and one GOE, respectively. The full line results from a fit of the analytical result for the $\Sigma^2$ statistics of the model [Eq.(\ref{eq:mixedH})] to the experimental one. This yielded $\lambda\leq 0.03$, $\lambda =0.13$ and $\lambda=0.20$ from top to bottom.  Reprinted from Phys. Rev. Lett. {\bf 81}, 4847 (1998).}
\label{sigma2gekop}
\end{figure}

In order to determine the symmetry-breaking parameter for the different experimental realizations of the coupling strength we, firstly, fitted as outlined in Ref.~\cite{Alt1998} the analytical result~\cite{Guhr1990a,Leitner1993} for the number variance $\Sigma^2$~\cite{Mehta1990} of the eigenvalues of the random matrices given in Eq.~(\ref{eq:mixedH}) to that of the unfolded resonance frequencies. In Fig.~\ref{sigma2gekop} the experimental data points with error bars and the analytical results (full lines) are shown for three different couplings together with those for the limiting cases of no symmetry, i.e., for $1$ GOE (dashed line), and of symmetry conservation, i.e., $2$ GOEs (dotted line).

\begin{figure}[h!]
\includegraphics[width=\linewidth]{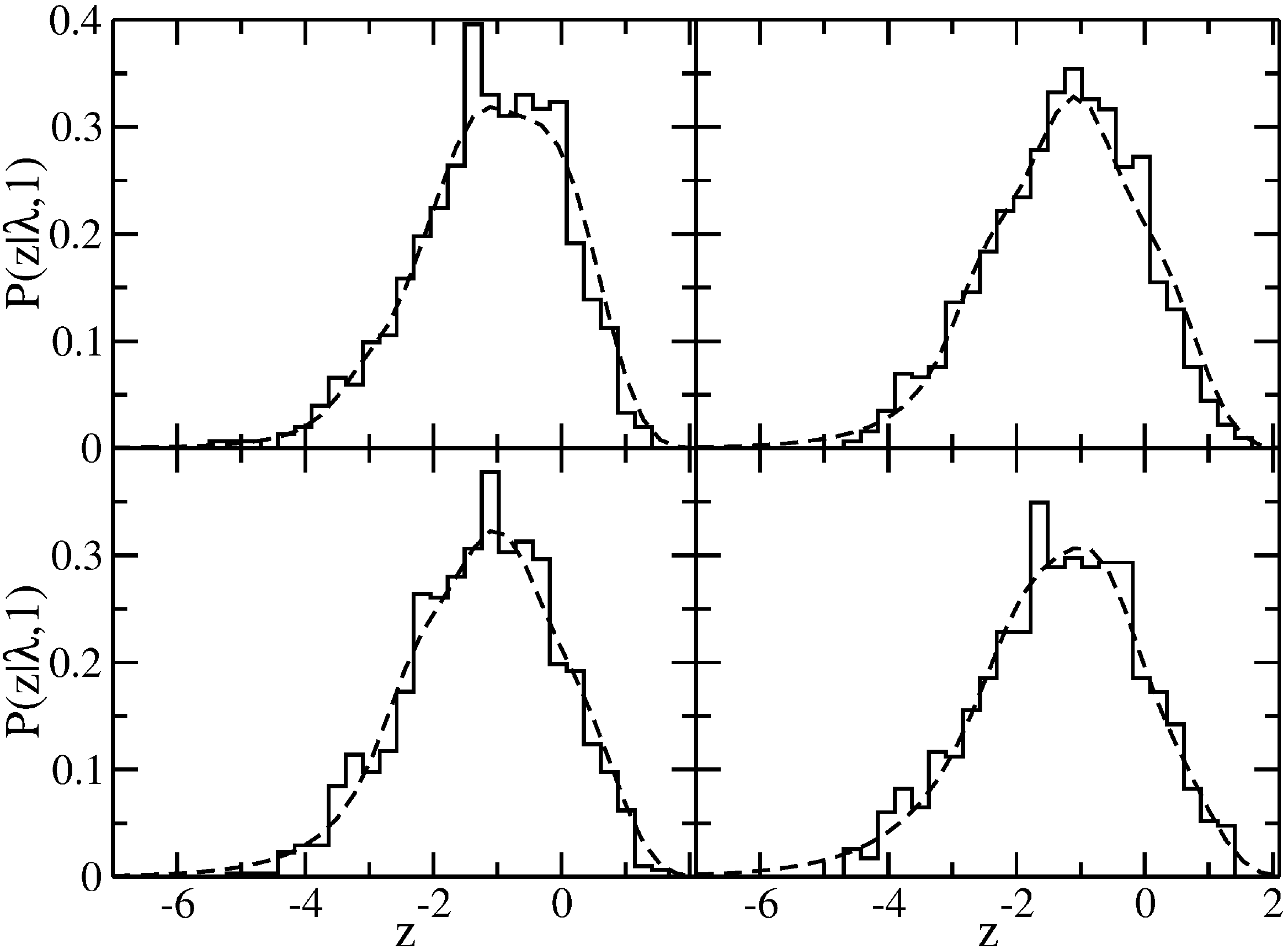}
\caption{Experimental strength distributions obtained from the resonances of the coupled stadium billiards for different couplings (histograms). The distributions of the analytical model derived from Eq.~(\ref{eq:mixedH}) including the threshold of detection, best fitting the experimental results are shown as dashed lines. The symmetry breaking parameters deduced from the fits equal $\lambda=0.003$, $0.09$, $0.14$ and $0.20$ from the upper left to the lower right panel.}
\label{strengthgekop}
\end{figure}

Secondly, in Ref.~\cite{Dietz2006a}, we compared the distributions of the resonance strengths $\Gamma_{\mu a}\Gamma_{\mu b}$ obtained from the measured resonances with a qualitative model developed based on the eigenvectors of the random matrices in Eq.~(\ref{eq:mixedH}). Figure~\ref{strengthgekop} shows the experimental (histograms) and analytical (dashed lines) results for one antenna combination and for four different couplings. Here, we took into account the experimental threshold of detection for narrow resonances and resonances with small strengths~\cite{Dembowski2005}. 
The symmetry-breaking parameters $\lambda$ obtained from both procedures agree within the experimental errors. The symmetry breaking actually turned out to be relatively weak even in the case of the strongest coupling, but nevertheless to be of the same size as the isospin symmetry breaking in $^{26}$Al, which yielded for the spreading width $\Gamma^\downarrow$ with respect to the mean level spacing, $\Gamma^\downarrow /D\simeq 0.25$, i.e., approximately every fourth resonance is influenced by the coupling.    
\section{\label{3d} Wave-dynamical chaos in three-dimensional resonators}
\begin{figure}[h!]
\includegraphics[width=0.6\linewidth]{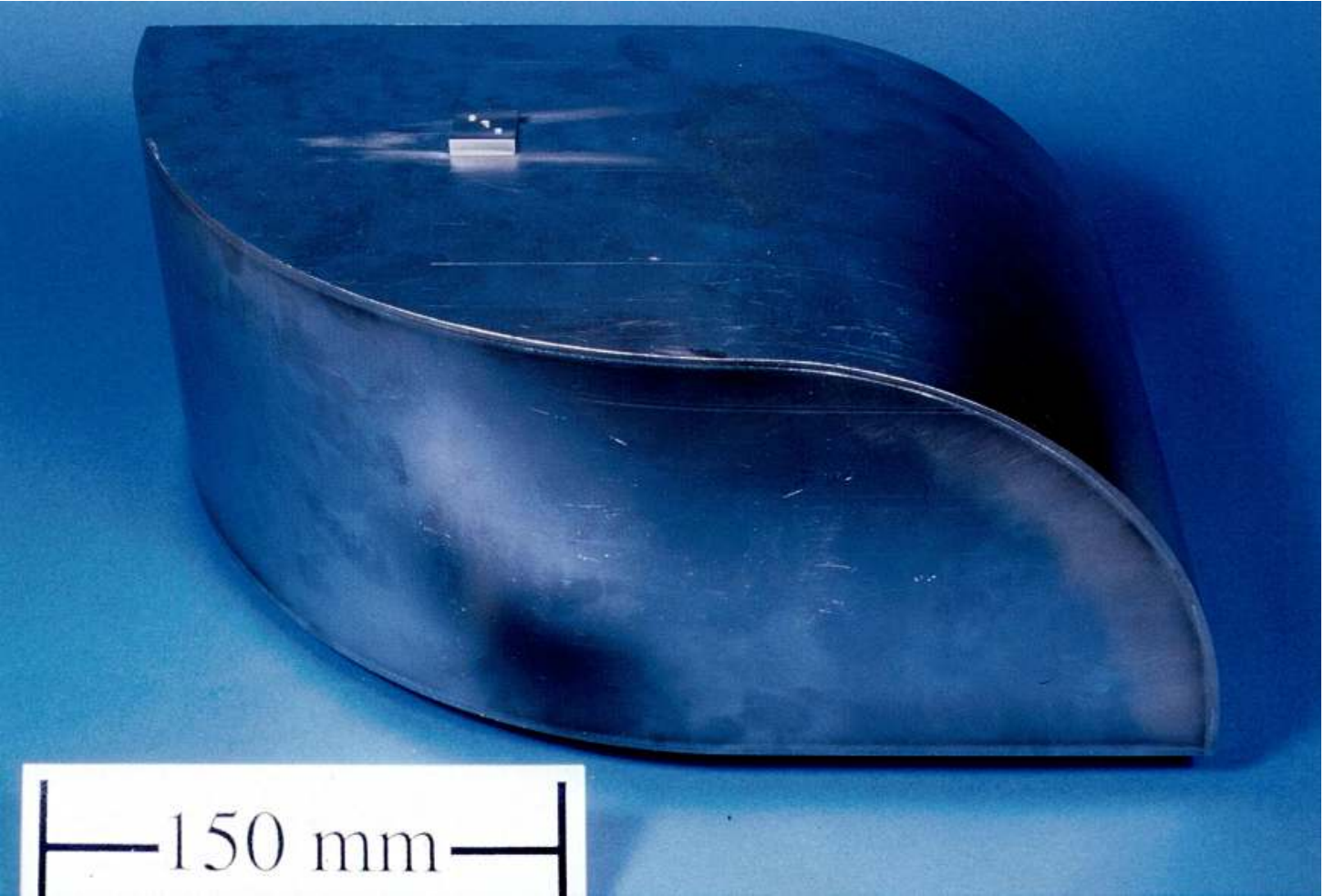}
\caption{(Color online) Photograph of the resonator with the shape of a three-dimensional desymmetrized stadium billiard. It was constructed from niobium.}
\label{photos3d}
\end{figure}

Another important issue besides the spectral fluctuation properties of a quantum system within the field of quantum chaos concerns the semiclassical approach. In this context, one of the most important achievements was the periodic orbit theory (POT) pioneered by Gutzwiller~\cite{Gutzwiller1990}, which paved the way for an understanding of the impact of the classical dynamics on the properties of the corresponding quantum system in terms of purely classical quantities. In contrast, RMT addresses exclusively system independent, i.e., universal properties. Yet, only recently, the “Bohigas-Giannoni-Schmit” conjecture~\cite{Bohigas1984} was proven on the basis of Gutzwiller's trace formula, which provides a semiclassical approximation for the fluctuating part of the resonance density of a quantum system in terms of a sum over the POs of the corresponding classical system~\cite{Heusler2007}. We have performed experimental tests of Gutzwiller's trace formula and its analogue for systems of chaotic, mixed or regular dynamics with metallic and dielectric resonators~\cite{Graf1992,Dembowski2001,Bittner2010a,Bittner2012a,Bittner2012b}. We also tested a Gutzwiller-like trace formula derived by Balian and Duplantier for three-dimensional electromagnetic resonators~\cite{Alt1996,Alt1997,Dembowski2002}. Here the main difficulty to overcome was the design of a microwave resonator that is completely chaotic and permits only a few non-generic modes. They were even present in our latest experiments with the resonator shown in Fig.~\ref{photos3d} which had the shape of a three-dimensional stadium billiard.
Note, that there is no analogy between three-dimensional microwave resonators and quantum billiards. Still, the former are most suitable for the study of wave-dynamical phenomena in chaotic systems. 

\begin{figure}[h!]
\includegraphics[width=\linewidth]{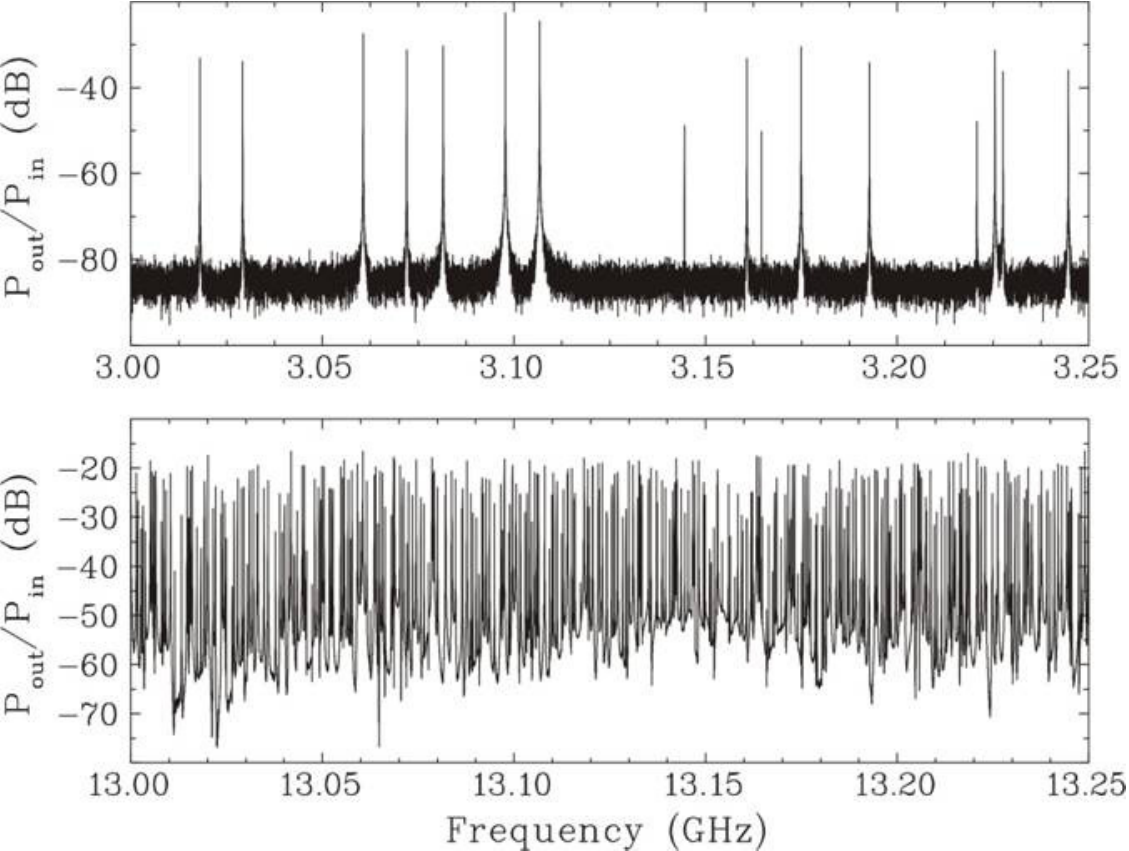}
\caption{Transmission spectrum of the three-dimensional resonator shown in Fig.~\ref{photos3d} for two different frequency ranges. Due to the high quality factors the resonances were well isolated up to $20$~GHz yielding a complete sequence of altogether 18764 resonance frequencies.}
\label{spektrum3dstadion}
\end{figure}

The three-dimensional desymmetrized stadium billiard consisted of two quarter cylinders with radii $r_1=200.0$~mm and $r_2=r_1/\sqrt{2}$~mm that were rotated by $90^\circ$ with respect to each other so that the height of each cylinder coincided with the radius of the other one. The resonator was constructed from niobium and the measurements were performed at liquid helium temperature, yielding exceptionally high quality factors of up to $10^7$. Consequently, a complete sequence of $18764$ resonance frequencies could be identified for excitation frequencies of up to $20$~GHz. Parts of one of the altogether $6$ measured transmission spectra are shown in Fig.~\ref{spektrum3dstadion}.   

\begin{figure}[h!]
\includegraphics[width=0.8\linewidth]{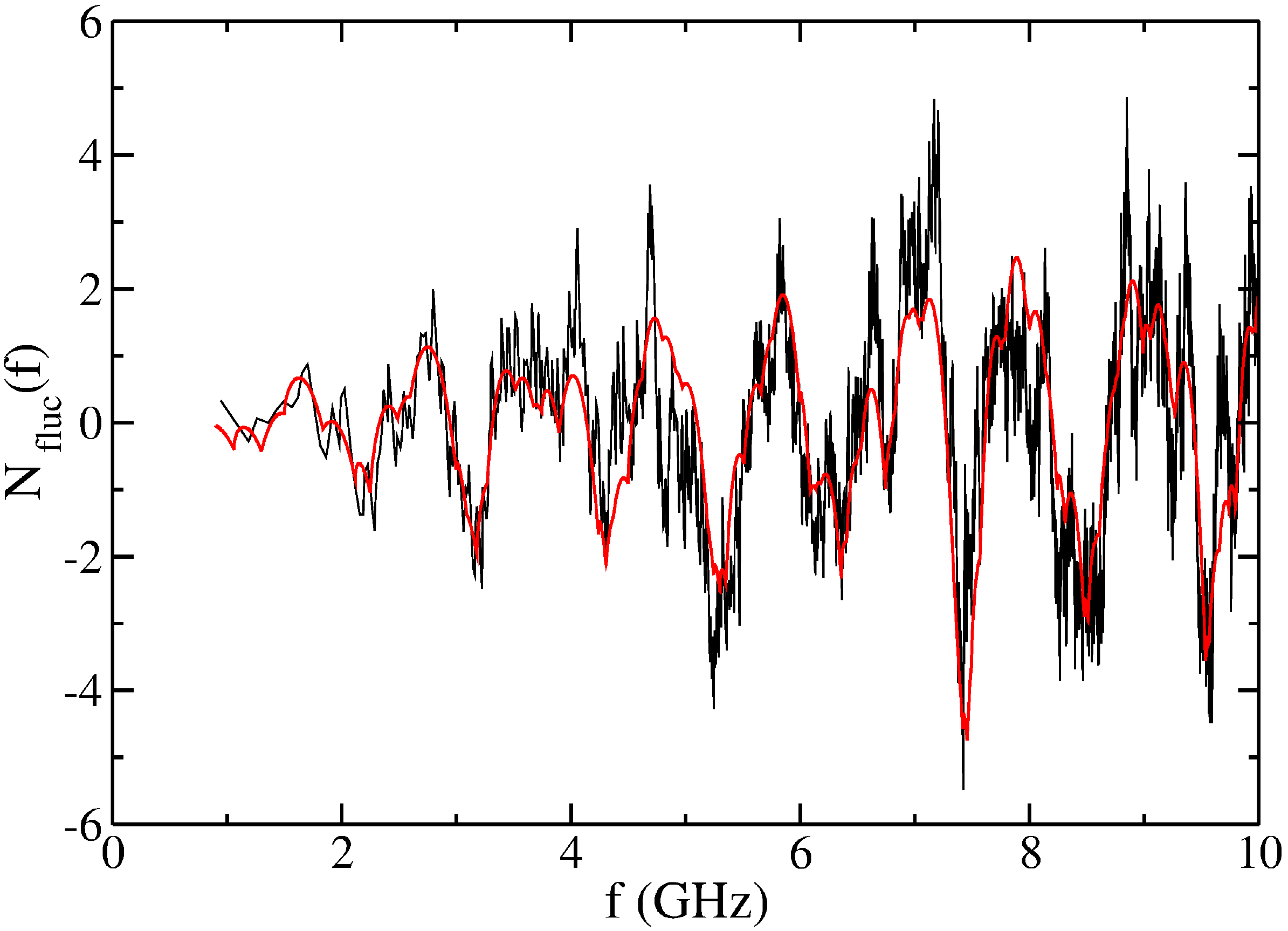}
\caption{(Color online) Fluctuating part of the integrated resonance density (black line) obtained for the resonator shown in Fig.~\ref{photos3d} compared to the contribution of the non-generic orbits (red [gray] line). Reprinted from Phys. Rev. Lett. {\bf 89}, 064101 (2002).}
\label{nfluc3dstadion}
\end{figure}

For the investigation of the fluctuation properties of the resonance frequencies these had to be unfolded with Weyl's formula for the smooth part $N_{smooth}(f)$ of the integrated resonance density $N(f)$ of electromagnetic cavities~\cite{Lukosz1973,Balian1977}. It is a polynomial of third order in the frequency, where, in distinction to that for the corresponding quantum billiard, the quadratic term is absent. The fluctuating part, $N_{fluc}(f)=N(f)-N_{smooth}(f)$, is shown in Fig.~\ref{nfluc3dstadion}. Like in the case of the two-dimensional stadium billiard in Fig.~\ref{nfluc2dstadium}, it exhibits slow oscillations, that again are due to non-generic POs. These consist of marginally stable bouncing-ball orbits with lengths $2r_1$ and $2r_2$ that bounce back and forth between the top and bottom walls of the quarter cylinders. Furthermore there are orbits in their common rectangular plane that are linearly stable within the plane and unstable with respect to perpendicular disturbances. An analytical expression for the contribution of the former to the resonance density was derived in~\cite{Alt1996}, that for the entire non-generic one in~\cite{Dembowski2002} on the basis of a quantum adiabatic method. The latter is plotted as red (gray) line in Fig.~\ref{nfluc3dstadion}.    

\begin{figure}[h!]
\includegraphics[width=0.8\linewidth]{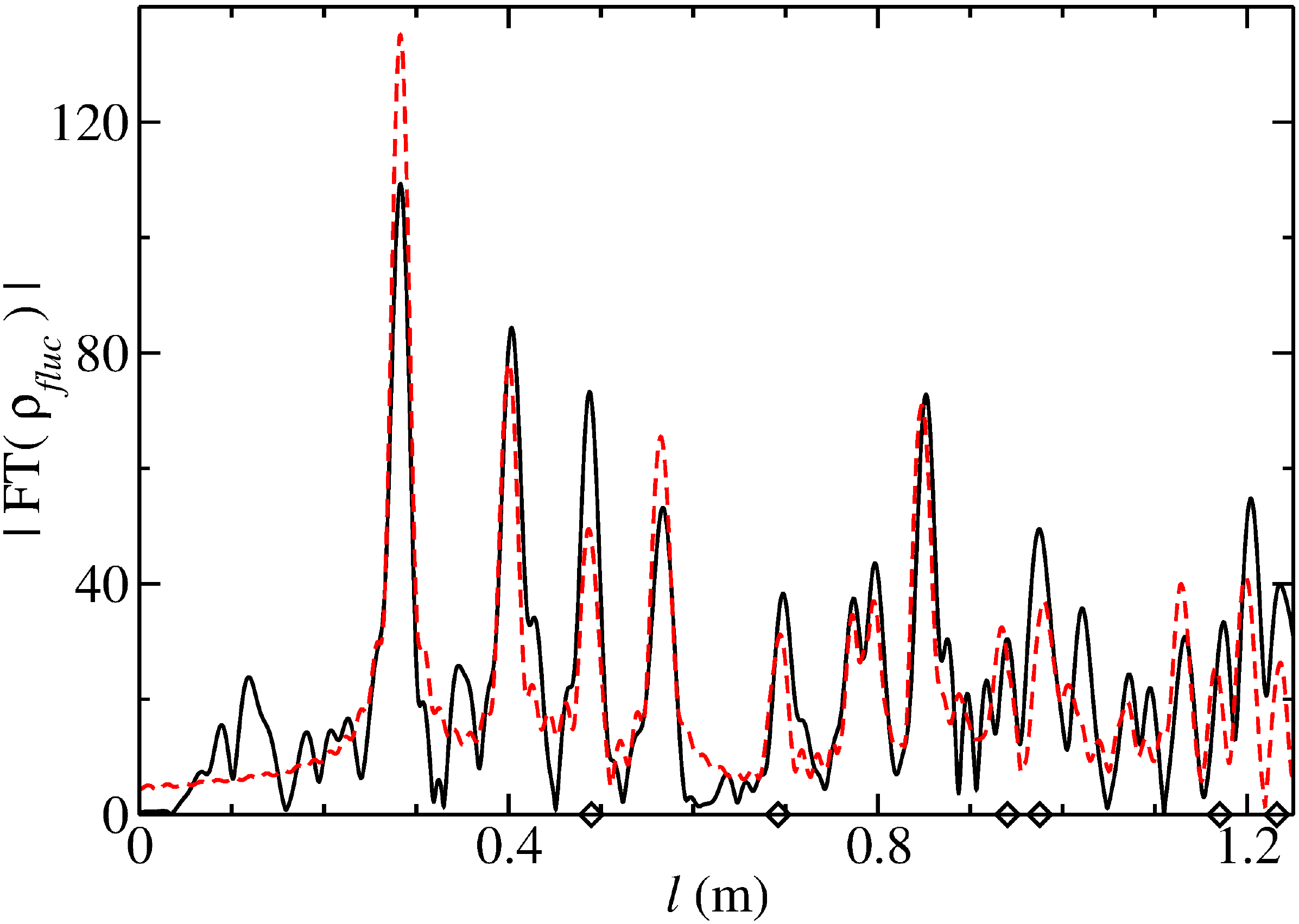}
\caption{(Color online) Fourier transforms of the fluctuating parts of the resonance densities obtained for the resonator shown in Fig.~\ref{photos3d} (black full line) and for the non-generic modes~\cite{Dembowski2002,Dietz2008b} plus the trace formula Eq.~(\ref{BD}) of Balian and Duplantier (red [gray] dashed line). Both curves exhibit peaks at the lengths of the POs. The diamonds mark the lengths of the POs in the common rectangular plane of the quarter cylinders.}
\label{fft3dstad}
\end{figure}

Figure~\ref{fft3dstad} exhibits the experimental length spectrum, deduced from the Fourier transform of the fluctuating part of the density of the resonance frequencies as black full line. The red dashed one was obtained from the sum of the Fourier transforms of the analytical results for the resonance density of the non-generic POs and the unstable ones. The latter is provided by the semiclassical trace formula of Balian and Duplantier~\cite{Balian1977},
\begin{equation}
\rho_{\rm fluc}(k)={1\over\pi}\sum_{p}{L_p\, 2\cos{\phi_p}\over
|{\rm det}(1-M_p)|^{1/2}}
\cos{\left(kL_p-{\pi\over 2}\mu_p\right)}.
\label{BD}
\end{equation}
The factor $2\cos{\phi_p}$ accounts for the polarization, that is, the vectorial character of the Helmholtz equation for three-dimensional resonators. It vanishes for orbits with an odd number of reflections. Note, that for separable systems, where the transversal electric and transversal magnetic modes decouple completely, $\vert 2\cos{\phi_p}\vert=2$. The quantities $L_p$ and $M_p$ denote the length of the periodic orbit $p$, and its stability matrix, respectively. They are identical with those entering Gutzwiller's trace formula~\cite{Gutzwiller1990}. The diamonds on the abscissa in Fig.~\ref{fft3dstad} mark the peaks that are exclucively assigned to POs in the common plane of the quarter cylinders. The other peaks below $l\simeq 0.6$~m are at the lengths of the bouncing-ball orbits. These indeed are comparatively large. 

\begin{figure}[h!]
\includegraphics[width=0.8\linewidth]{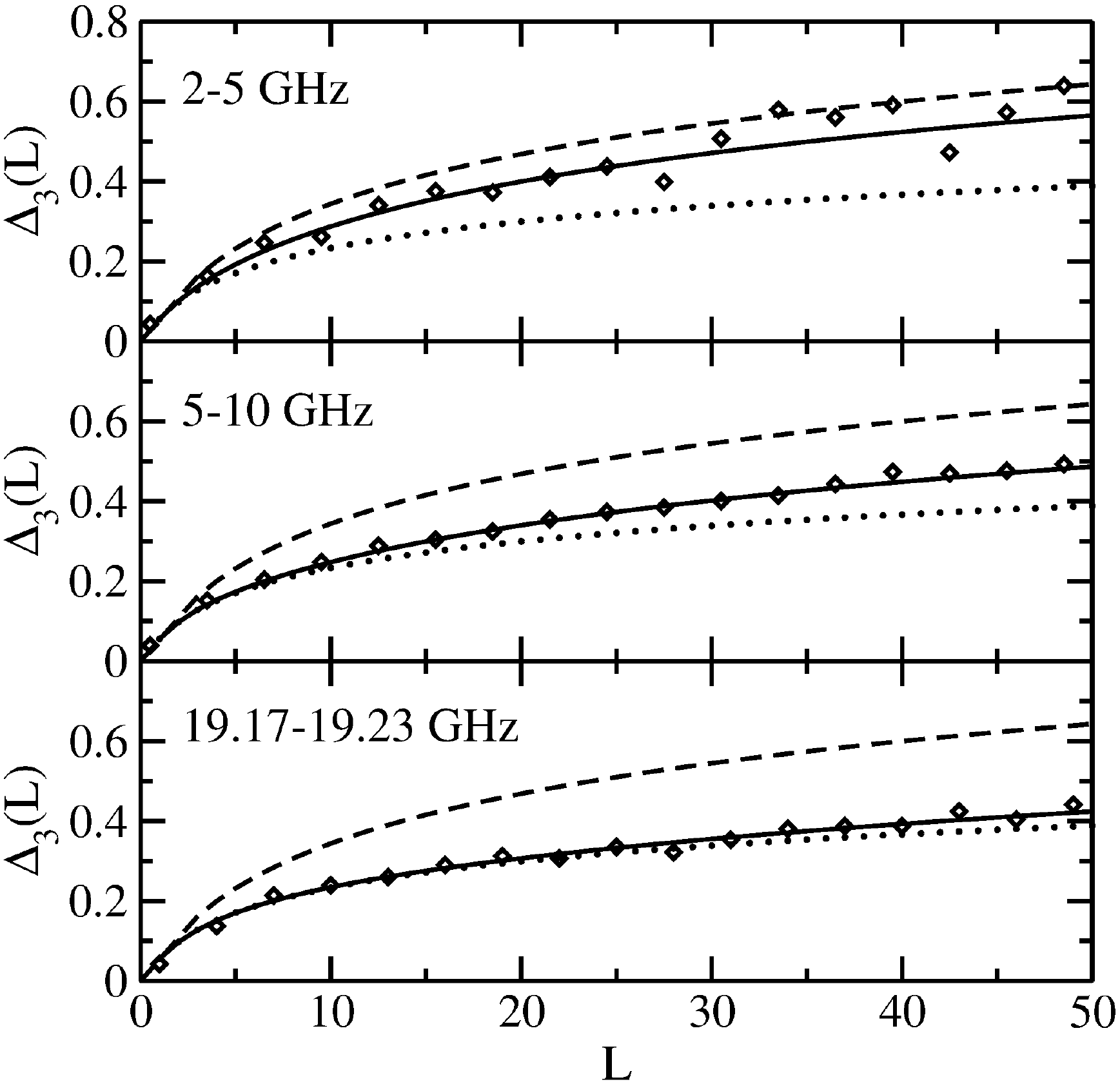}
\caption{The $\Delta_3$ statistics of the resonance frequencies of the resonator shown in Fig.~\ref{photos3d} evaluated in the three frequency intervals indicated in the insets (diamonds). The dotted curve corresponds to that for one GOE, the dashed one to that of two uncoupled GOEs, i.e., to the case $\lambda =0$ in Eq.~(\ref{eq:mixedH}), and the full one to the curve deduced from Eq.~(\ref{eq:mixedH}) best fitting the experimental one yielding $\lambda=0.005,\, 0.124,\,  0.498$ from top to bottom.}
\label{delta33dstadion}
\end{figure}

In conclusion, the contributions from non-generic orbits could not be neglected and had to be extracted for the study of the spectral fluctuation properties of the unfolded resonance frequencies.    
These were expected to coincide with the RMT result for the GOE, since the dynamics of the corresponding classical billiard was shown to be chaotic~\cite{Papenbrock2000,Bunimovich2006}. In Fig.~\ref{delta33dstadion} the experimental result for the $\Delta_3$ statistics is plotted (diamonds) for three frequency ranges. Even though the non-generic contributions were extracted by proceeding as in the case of the two-dimensional stadium billiard (see ~\refsec{2dstadium}), we observe deviations from the GOE result (dotted lines). Instead, the $\Delta_3$ statistics is very well decribed by that of random matrices of the form in Eq.~(\ref{eq:mixedH}), applicable to sytems consisting of two coupled chaotic ones (full lines). Actually, in the frequency range of $2-5$~GHz it is close to that of two uncoupled GOEs (dashed line). 

The reason for the deviations from the GOE results is not yet fully understood. Numerical studies~\cite{Dietz2008b} of the spectral properties of three-dimensional quantum stadium billiards demonstrated that they cannot be attributed to the partial breaking of the reflection symmetry present, when the radii $r_1$ and $r_2$ of the two quarter cylinders are equal. Furthermore, our interpretation of the good agreement of the spectral properties with those of the model in Eq.~(\ref{eq:mixedH}) as a partial decoupling of the electric and the magnetic field modes could not yet be confirmed, neither experimentally nor numerically. However, the fact that a considerable fraction of the POs have $\vert 2\cos{\phi_p}\vert=2$ provides a supportive indicator for its correctness. In the following section we will present another system whose spectral properties deviate from the RMT predictions.   
\section{\label{triangular} Spectral fluctuations of a system with threefold symmetry}
\begin{figure}[h!]
\includegraphics[width=0.5\linewidth]{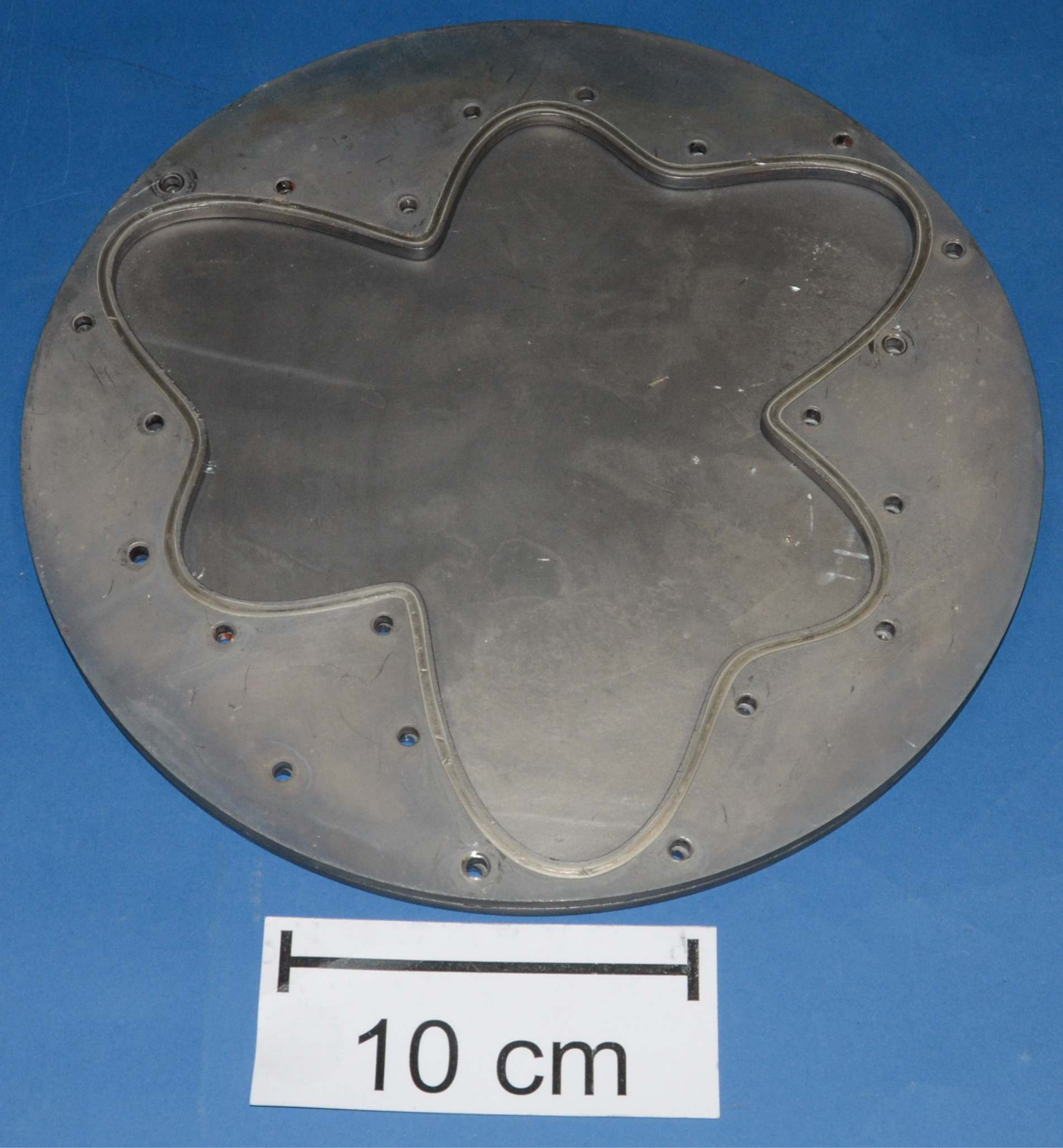}
\caption{(Color online) Photograph of the microwave billiard with threefold symmetry and no further symmetries. It consisted of a bottom plate, a middle plate, which had a hole with the shape of the billiard, and a lid. All parts were constructed from copper and covered with lead before they were squeezed together tightly with screws along the boundary of the cavity. The top plate is removed.}
\label{phototriangular}
\end{figure}

As has been already stated above, according to the “Bohigas-Giannoni-Schmit” conjecture~\cite{Bohigas1984} the spectral properties of time-reversal (\cT) invariant chaotic systems coincide with those of random matrices drawn from the GOE. This, however, is not true for systems with an unidirectional classical dynamics (see~\refsec{tunneling}) or for billiards that have a shape with a threefold symmetry and no further one like, e.g., a mirror symmetry~\cite{Leyvraz1996}. In this context, we investigated the spectral properties of the microwave billiard presented in Fig.~\ref{phototriangular}. Its shape is invariant under rotations of $120^\circ$. 
The eigenfunctions $\Psi$ of the Hamiltonian of such a quantum billiard can be classified according to their transformation properties with respect to a rotation $R$ by $120^\circ$, $R\Psi =\exp\left(2\pi il/3\right)\Psi ,\, l=-1,0,1$. Due to \cT invariance the eigenvalues associated with the classes $l=\pm 1$ are degenerate. Accordingly, the eigenvalue spectrum is composed of singlets corresponding to the case $l=0$ and doublets of identical ones~\cite{Leyvraz1996,Dietz2005}. The spectral properties of the singlets and the doublets have been predicted to coincide with those of random matrices from the GOE and the GUE, respectively~\cite{Leyvraz1996}. 

\begin{figure}[h!]
\includegraphics[width=0.6\linewidth]{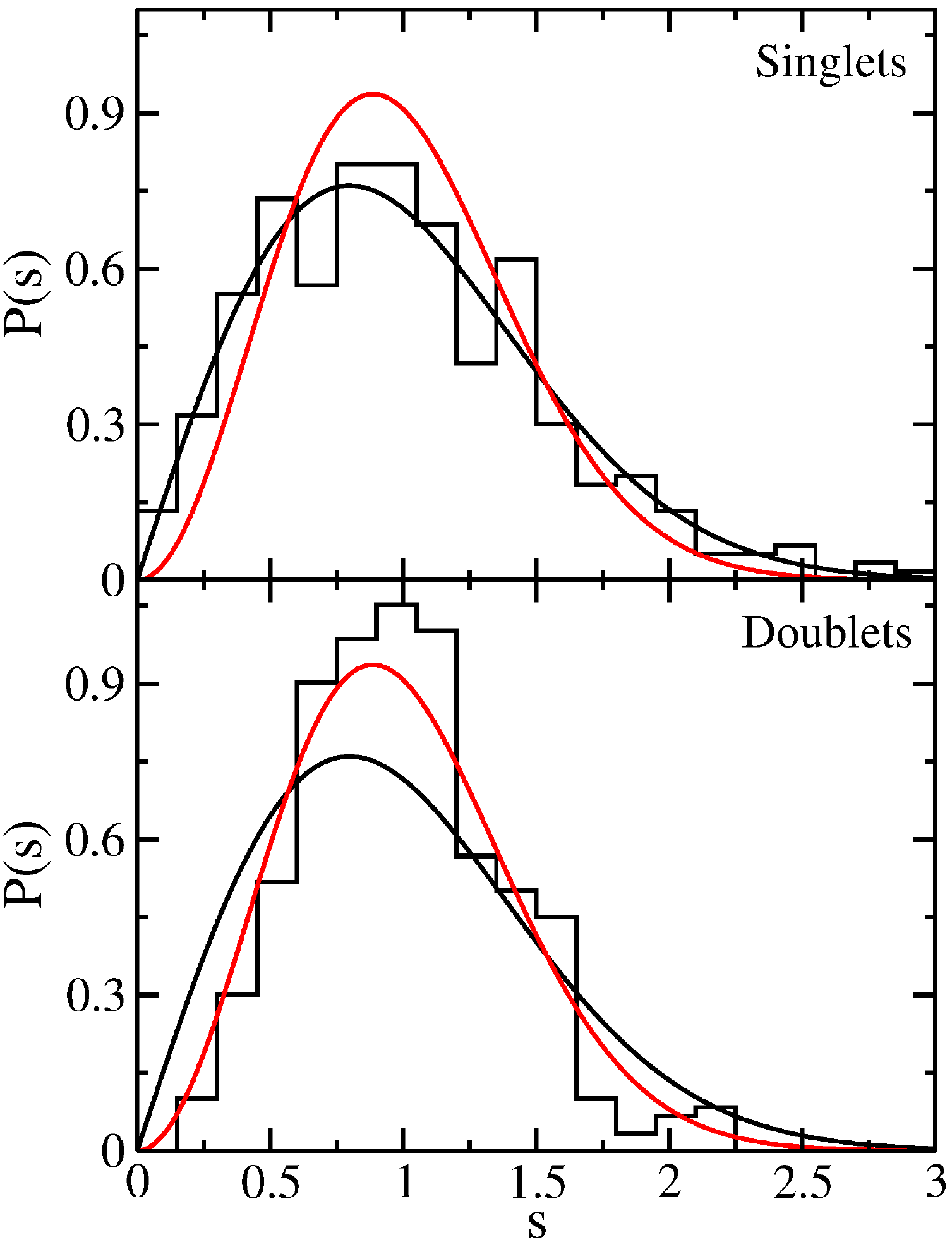}
\caption{(Color online) Nearest-neighbor spacing distributions of the resonance frequencies of the singlets (upper panel) and the doublets (lower panel) obtained for the microwave billiard shown in Fig.~\ref{phototriangular} (histogram). The RMT results for the GOE and the GUE are shown as black and red (gray) lines, respectively.}
\label{absttriangular}
\end{figure}

\begin{figure}[h!]
\includegraphics[width=0.6\linewidth]{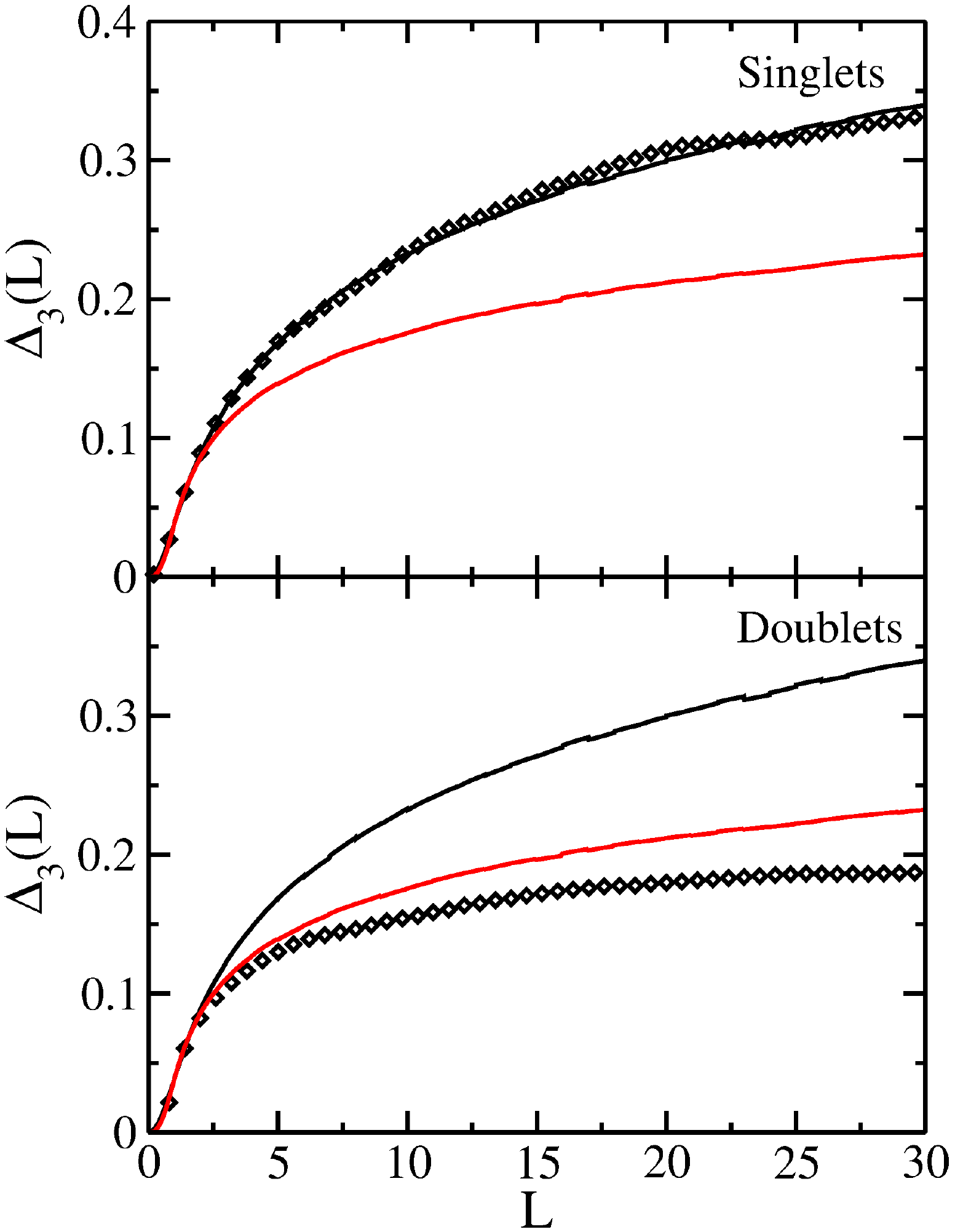}
\caption{(Color online) The $\Delta_3$ statistics of the resonance frequencies of the singlets and the doublets obtained for the microwave billiard shown in Fig.~\ref{phototriangular} (diamonds). The RMT results for the GOE and the GUE are plotted as black and as red (gray) lines, respectively.  Reprinted from Phys. Rev. Lett. {\bf 90}, 014102 (2003).}
\label{delta3triangular}
\end{figure}

We determined the eigenvalues of the billiard with the shape shown in Fig.~\ref{phototriangular} by performing measurements at superconducting conditions. Still, since the degeneracy of the doublet states was only partly split due to mechanical imperfection~\cite{Dembowski2000} a more sophisticated method had to be developed for their identification~\cite{Dembowski2003a}. It utilized the dependence of the behavior of the eigenmodes under the variation of the relative phase between two input antennas on their symmetry properties. In Fig.~\ref{absttriangular} the nearest-neighbor spacing distributions of the singlets and doublets (histograms) are compared with the RMT result for the GOE (black line) and for the GUE (red [gray] line), respectively. Similarly, the $\Delta_3$ statistics of the singlets and doublets (diamonds) is presented in Fig.~\ref{delta3triangular} together with the GOE (black line) and the GUE (red [gray] line), respectively. These figures demonstrate that the experimental results were in accordance with the theory.  
\section{\label{tunneling} Dynamical tunneling in constant-width billiards}
\begin{figure}[h!]
\includegraphics[width=0.5\linewidth]{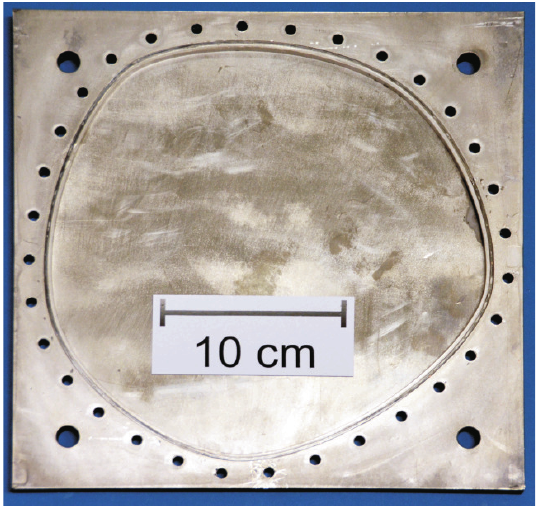}
\caption{(Color online) Photograph of the microwave billiard with the shape of the constant-width billiard B1. It was constructed from copper plates. The cavity was milled out of the bottom plate. The top plate is removed.  Reprinted from Phys. Rev. E {\bf 90}, 022903 (2014).}
\label{photocw}
\end{figure}

\begin{figure}[h!]
\includegraphics[width=0.8\linewidth]{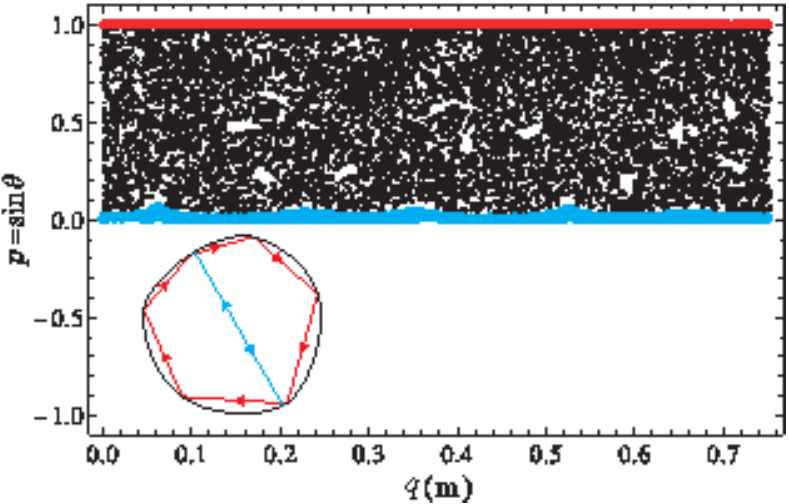}
\caption{(Color online) Poincar\'e surface of section (PSOS) for billiard B1 where all particles were launched into the billiard in the clockwise direction. Due to unidirectionality only the upper half of the PSOS is filled. The chaotic sea is bordered by a region of whispering gallery modes around $p=1$ plotted in red (gray) and one of orbits around the diameter orbit plotted in blue (light gray). The inset shows examples for the associated orbits.  Reprinted from Phys. Rev. E {\bf 90}, 022903 (2014).}
\label{psscw}
\end{figure}

Recently, we investigated the phenomenon of dynamical tunneling in two microwave billiards with the shapes of constant-width billiards~\cite{Knill1998,Gutkin2007}. A photograph of one is shown in Fig.~\ref{photocw}. 
Such billiards possess the particular property of unidirectionality, that is, when a particle is launched into the billiard with a certain rotational direction then it will continue to rotate in that direction forever. Accordingly, when constructing a Poincar{\'e} surface of section (PSOS) by launching many particles into the billiard in clockwise direction, then only its upper half will be filled, as illustrated in Fig.~\ref{psscw}. Here, $p=\sin\theta$ where $\theta$ is the angle between the particle trajectory and the normal to the boundary at the point of impact, and $q$ is the location of the latter in terms of the arclength along the boundary. The PSOS consists of a large chaotic sea and two regions of regular KAM tori. Firstly, there is a barrier region around the diameter orbit at $p=0$, shown in blue (light gray). Secondly, there is a region of whispering gallery modes around $p=1$.  Examples of both types of orbits are plotted in the inset in the corresponding colors. Furthermore, we found very tiny regular islands inside the chaotic sea for one of the two investigated billiards~\cite{Dietz2014}. 

The fact, that only one half of the PSOS is filled is a clear indication that classically the transition from clockwise to anticlockwise motion, i.e., the transition from positive to negative $p$ values is strightly forbidden. In contrast, in the corresponding quantum billiard this is possible via tunneling through the barrier region around $p=0$. This manifests itself in a partial lifting of the degeneracies of pairs of eigenenergies associated with the clockwise and anticlockwise quasimodes~\cite{Dietz2014}. Accordingly, the size of the tunneling effect is reflected in that of the splittings of the eigenenergies.    

The objective of the experiments with the two microwave billiards with the shapes of constant-width billiards, referred to as billiards B1 amd B2 in the sequel, was to investigate this phenomenon with unprecedented accuracy using superconducting microwave billiards. Both billiards had the same diameter $2r=48$~cm and their shapes were barely distinguishable, whereas the widths of the barrier and the whispering gallery orbit regions were considerably larger in the PSOS of B1 than in that of B2. More details concerning the measurements are provided in Ref.~\cite{Dietz2014}. 

\begin{figure}[h!]
\includegraphics[width=0.8\linewidth]{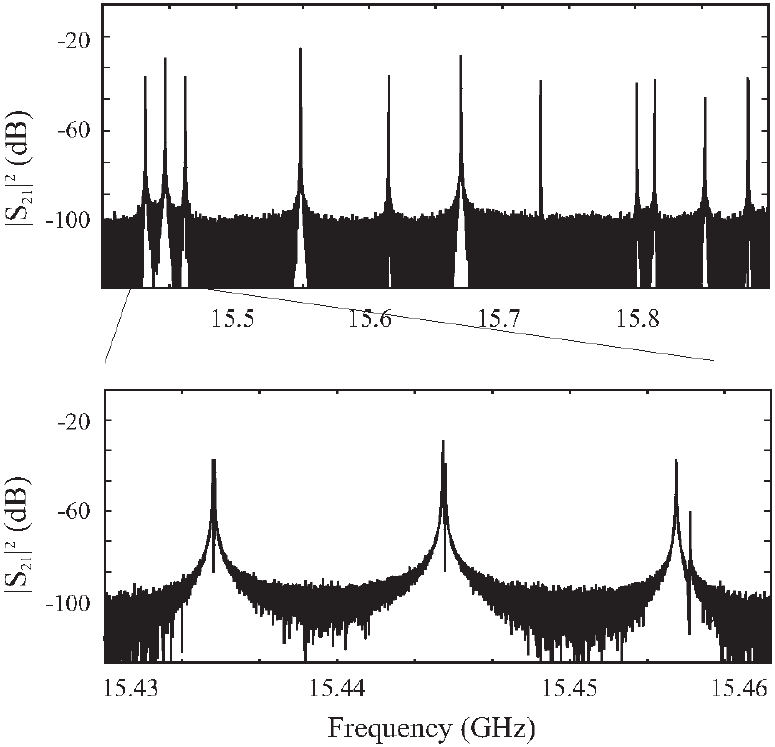}
\caption{Resonance spectrum of billiard B1. The upper panel shows it in a frequency range from 15.4 to 15.9~GHz, the lower one a zoom into a triplet of nearly degenerate pairs of resonances.  Reprinted from Phys. Rev. E {\bf 90}, 022903 (2014).}
\label{spectrumcw}
\end{figure}

To demonstrate the high precision of the measurements and also its indispensability for the identification of the doublets we show in Fig.~\ref{spectrumcw} two parts of the transmission spectrum of billiard B1.    
The resonances visible in the upper panel seem to be sharp and well isolated. However, a zoom into the region of the triplett observed at the left side of the spectrum, reveals that they all correspond to doublets of nearly degenerate resonance frequencies. In total, $~390$ doublets and $15$ singlets were detected in billiards B1 and B2, respectively. The latter correspond to diameter orbits, that are analogues of those of the circle billiard of the same diameter $2r$, the eigenvalues $k_n^c$ are the zeroes of the Bessel functions, $J_0(k_n^cr)=0$. We actually identified \emph{all} singlets in the list of resonance frequencies of B1 and B2 by comparison with these zero-momentum modes.

In order to check the completeness of the sequence of doublets and to study the spectral properties of the constant-width billiards, we separated the resonance frequencies into singlets, and the smaller and larger ones of the doublets, i.e., the left, $f_n^l$, and the right, $f_n^r$, doublet partners in the resonance spectra (see lower panel of Fig.~\ref{spectrumcw}), respectively. This was possible, because the correlations between the doublet partners and those between the singlets and the doublets were much weaker than those amongst the members of each of the sequences. Consequently, we could simply disregard the singlets and investigate the fluctuation properties of the left and the right doublet partners separately. For this, we first unfolded them with Weyl's formula~\cite{Weyl1912} (see~\refsec{2dstadium}). The comparison of the smooth part of the integrated density of the resonance frequencies with the latter and also the shape of its fluctuating part confirmed that we identified the \emph{complete} sequence of doublets. Due to the unidirectionality of the classical dynamics, the spectral properties were predicted to coincide with the RMT result for the GUE, even though the system is \cT invariant~\cite{Gutkin2007}. We didn't find any agreement with the GUE and also not with the GOE results, but instead found very good agreement with those of a random-matrix model for systems with violated \cT invariance and mixed dynamics. Within this model the form of the random matrices is given by Eq.~(\ref{eq:mixedH}), with $H_1$ replaced by a random matrix from the GUE, and $H_2$ by a diagonal matrix with random Poissonian numbers as entries. The elements of the matrix $V$, which describes the mixing of the chaotic and the regular states modeled by $H_1$ and $H_2$, respectively, were chosen as Gaussian-distributed random numbers. Thus, even though the classical dynamics of the investigated billiards was not fully chaotic, the predictions of Ref.~\cite{Gutkin2007} could be confirmed.
\begin{figure}[h!]
\includegraphics[width=0.6\linewidth]{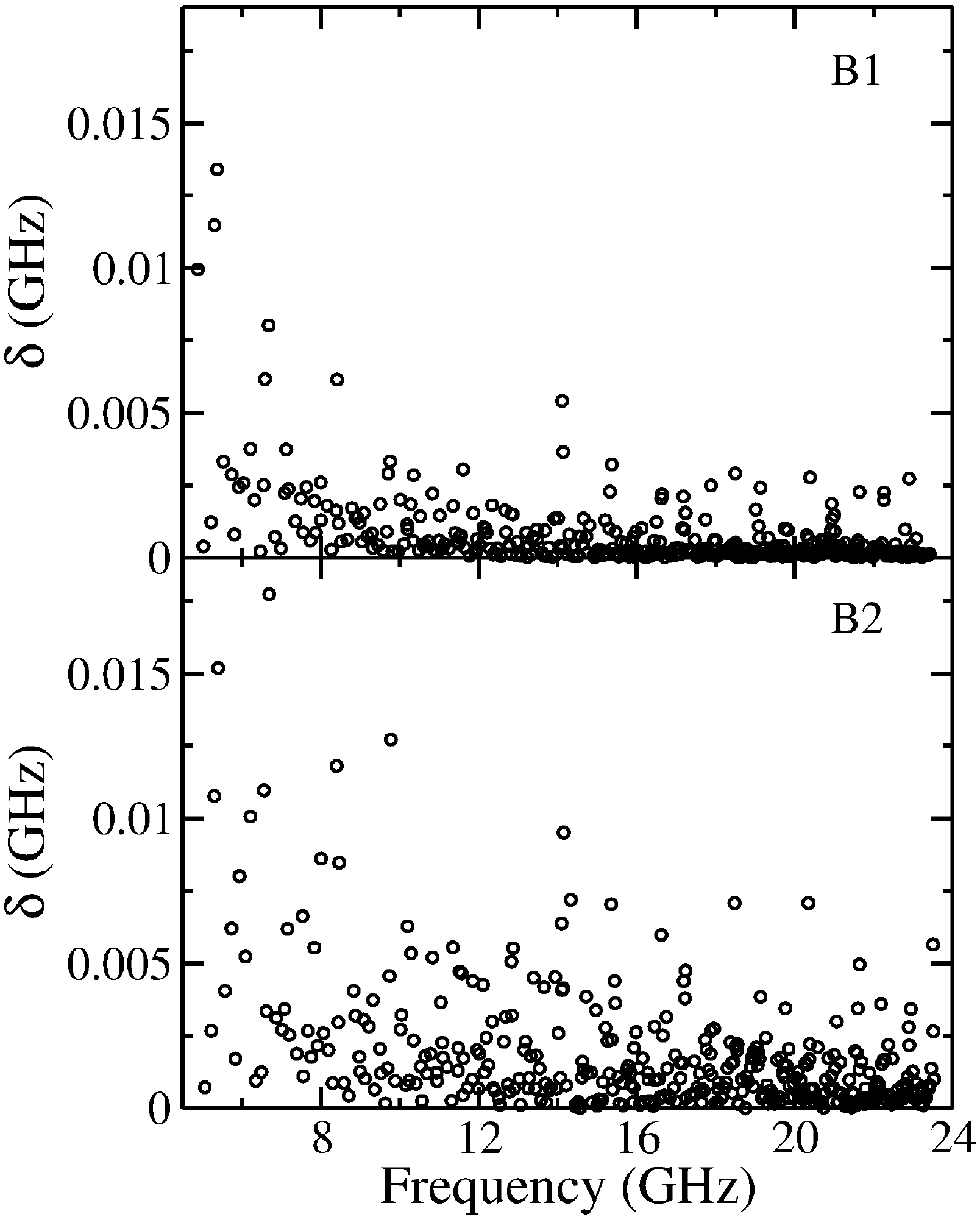}
\caption{(Color online) The splittings of the resonance frequencies of the doublets, $\delta=f_n^r-f_n^l$, for B1 (upper panel) and B2 (lower panel) versus their average resonance frequencies. They are by a factor of $10^2-10^3$ smaller than their average resonance frequency. This underlines again the necessity of high-precision experiments to identify a complete sequence of doublets.}
\label{splitcw}
\end{figure}

The splittings $\delta=f_n^r-f_n^l$ of the resonance frequencies of the doublet partners are plotted versus their average resonance frequency for B1 in the upper panel and for B2 in the lower one of Fig.~\ref{splitcw}, respectively. The typical splitting size is smaller for B1 than for B2. This is in agreement with the expectation, that the tunneling rate through the barrier region in the PSOS, which is broader for B1 than for B2, decreases with increasing width of that region. Furthermore, the average splitting size diminishes algebraically with increasing frequency, not exponentially as might be expected semiclassically. This can be attributed to the fact, that the PSOS consists of a chaotic region and several regular regions, where the splitting sizes diminish exponentially with different rates. We considered all splittings, i.e., the sum over the exponentials which yields an algebraical decay~\cite{Dittes2000}.
\begin{figure}[h!]
\includegraphics[width=0.7\linewidth]{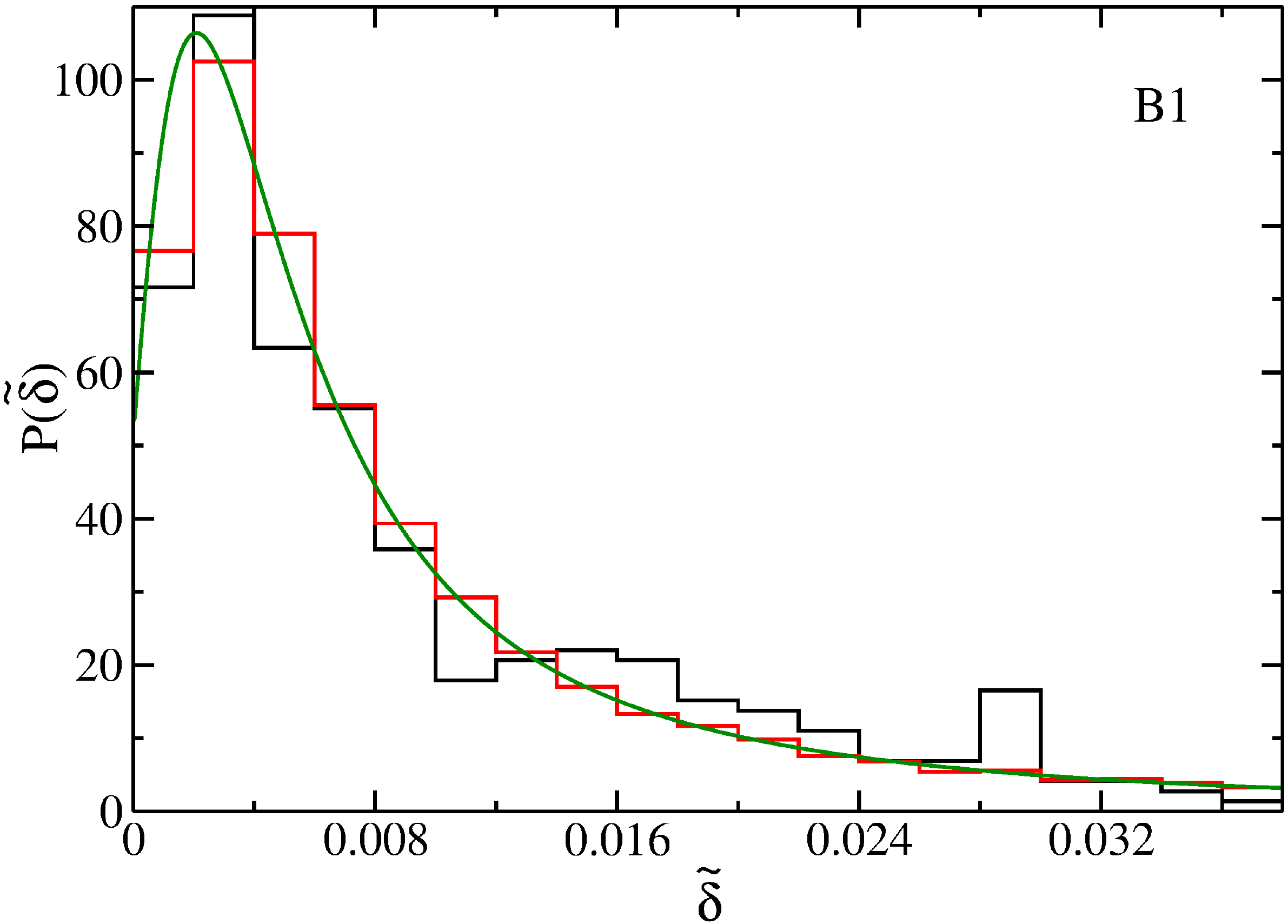}
\includegraphics[width=0.7\linewidth]{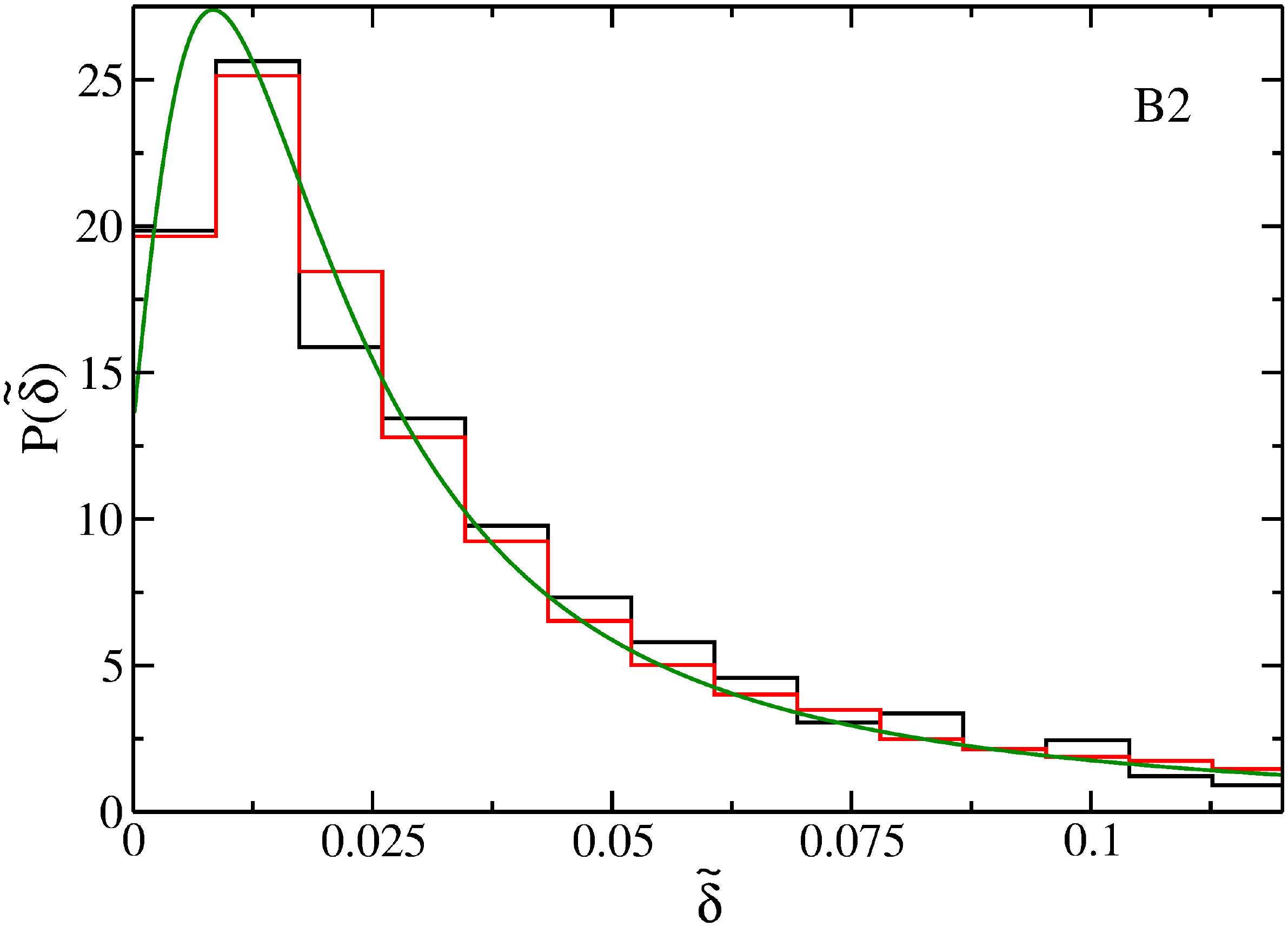}
\caption{(Color online) Comparison of the experimentally determined splitting distributions (black histograms) of B1 [upper panel] and B2 [lower panel] with that obtained from the random-matrix model of Eq.~(\ref{eq:1}) shown as red (gray) histograms and with the analytical result of Eq.~(\ref{spl_distr}) plotted as green (light gray) line. Here, the tunneling parameters deduced from the fit of the RMT model to the experimental distributions are $\tau=0.094$ for B1 and $\tau=0.185$ for B2. Reprinted from Phys. Rev. E {\bf 90}, 022903 (2014).}
\label{splitdistrcw}
\end{figure}

For a better understanding of the tunneling we investigated also the distribution of the splittings. The black histograms in the upper and lower panel in Fig.~\ref{splitdistrcw} correspond to the experimental results for billiards B1 and B2, respectively. In order to extract system specific quantities, like the area and the perimeter of the billiards, we used the splittings $\tilde\delta$ of the unfolded left and right doublet partners. We demonstrated in Ref.~\cite{Dietz2014} that the shapes of the splitting distributions are not affected by states that are associated with the regular parts of the PSOS, that is, the chaotic component dominates. They, indeed, are well described by a random-matrix model of the form
\begin{equation}
H^{splitting}=\left({\begin{array}{ccc}
        H\, &0\, &0\\
        0\, &D_0\, &0\\
        0\, &0\, & H^*\\
        \end{array}}
  \right)\,
+
\tau D\left({\begin{array}{ccc}
        0\, &V\, &0\\
        V^\dagger\, &0\, &V^\dagger\\
        0\, &V\, & 0\\
        \end{array}}
  \right)\, ,
\label{eq:1}
\end{equation}
which is applicable to systems with broken \cT invariance exhibiting tunneling between two regions with chaotic dynamics.
Here, $H$ and $H^*$ are $2\times 2$ random matrices from the GUE, that are associated with the parts of the Hilbert space corresponding to clockwise and anticlockwise motion, respectively, and $D_0$ is a random Poissonian number which represents a singlet state. The tunneling from the upper to the lower part of the PSOS via the barrier region, that is, the coupling of $H$ and $H^*$ via the singlet state, is caused by the off-diagonal matrix $V$, which has Gaussian-distributed random numbers as entries. The strength of the coupling is determined by the tunneling parameter $\tau$. It was obtained by fitting the splitting distribution of the RMT model of Eq.~(\ref{eq:1}) to the experimental ones. On the basis of the latter we furthermore derived an analytical expression for the splitting distribution, 
\begin{equation}
{\rm P}(\tilde\delta)=\frac{2}{\sqrt{\pi}}\frac{\tau^2}{(\tilde\delta+\tau^2)^2}e^{-4\frac{\tau^4}{(\tilde\delta+\tau^2)^2}}\left[1+8\frac{\tau^4}{(\tilde\delta+\tau^2)^2}\right],
\label{spl_distr}
\end{equation}
which we believe is generally applicable to systems with violated \cT invariance, where dynamical tunneling takes place between two regions with a chaotic dynamics. The green (light gray) lines in Fig.~\ref{splitdistrcw} correspond to the analyticial results computed for the parameter values of $\tau$ given in the captions. Its value is smaller for billiard B1 than for B2, as expected, because the PSOS of the former comprises a broader barrier region. The agreement between the experimental, the RMT and the analytical results for the splitting distributions is very good. Remarkably, the RMT model and the analytical formula for the splitting distribution depend only on one parameter which provides information on the tunneling rate.  

\begin{figure}[h!]
\includegraphics[width=\linewidth]{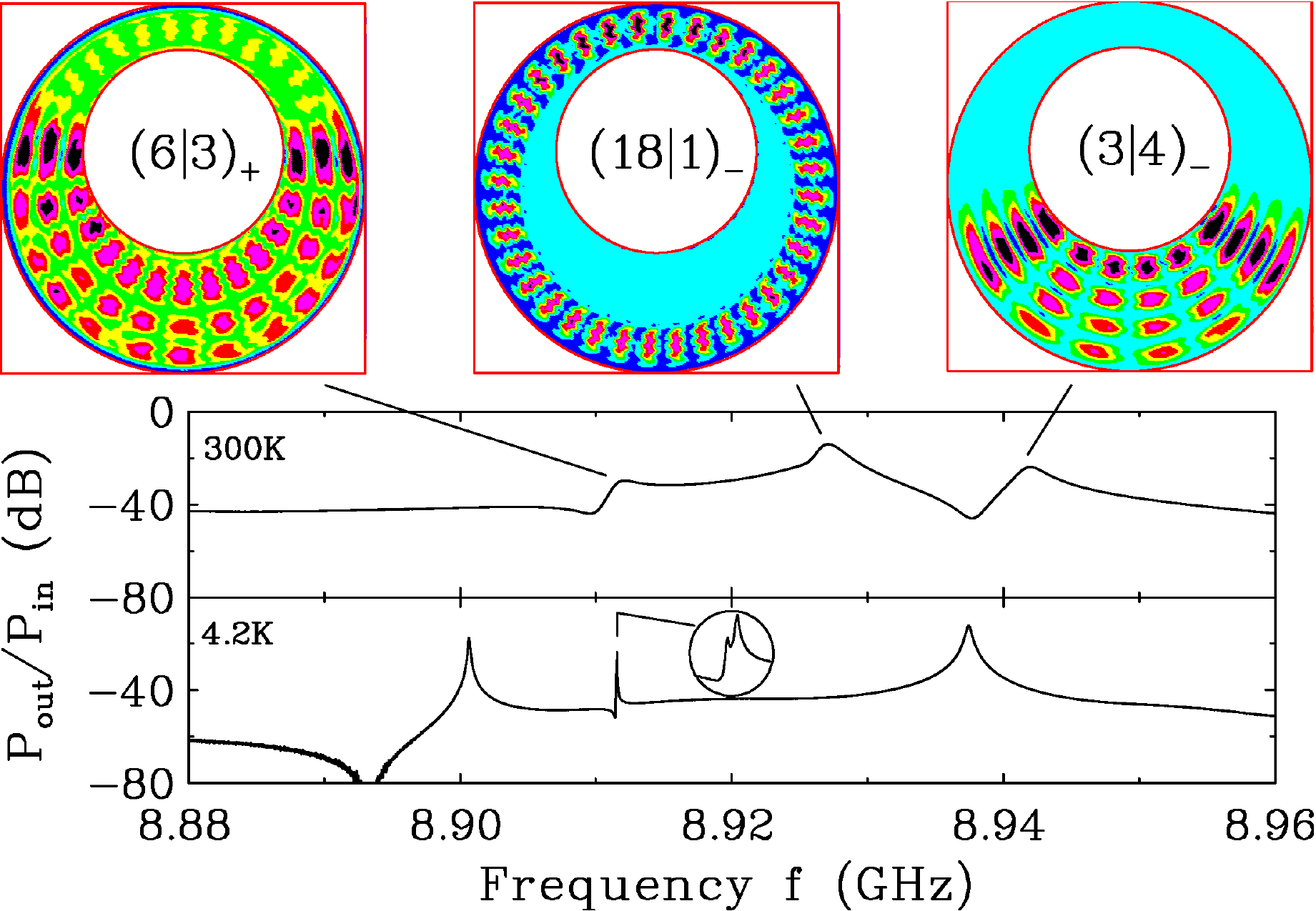}
\caption{(Color online) Transmission spectrum of the annular billiard shown schematically in the three top panels in the frequency range from $8.88$ to $8.96$~GHz obtained at room temperature (upper panel) and at 4.2 K (lower panel). The combined measurement of field distributions with the normal conducting microwave billiard and of resonance spectra with the superconducting one allows the identification of the quasidoublets and the assigment of quantum numbers $(m,n)$ and of a parity with respect to the symmetry axis like for the three modes associated with the resonances, shown in the top panels. Reprinted from Phys. Rev. Lett. {\bf 84}, 867 (2000).}
\label{splitannular}
\end{figure}

\begin{figure}[h!]
\includegraphics[width=\linewidth]{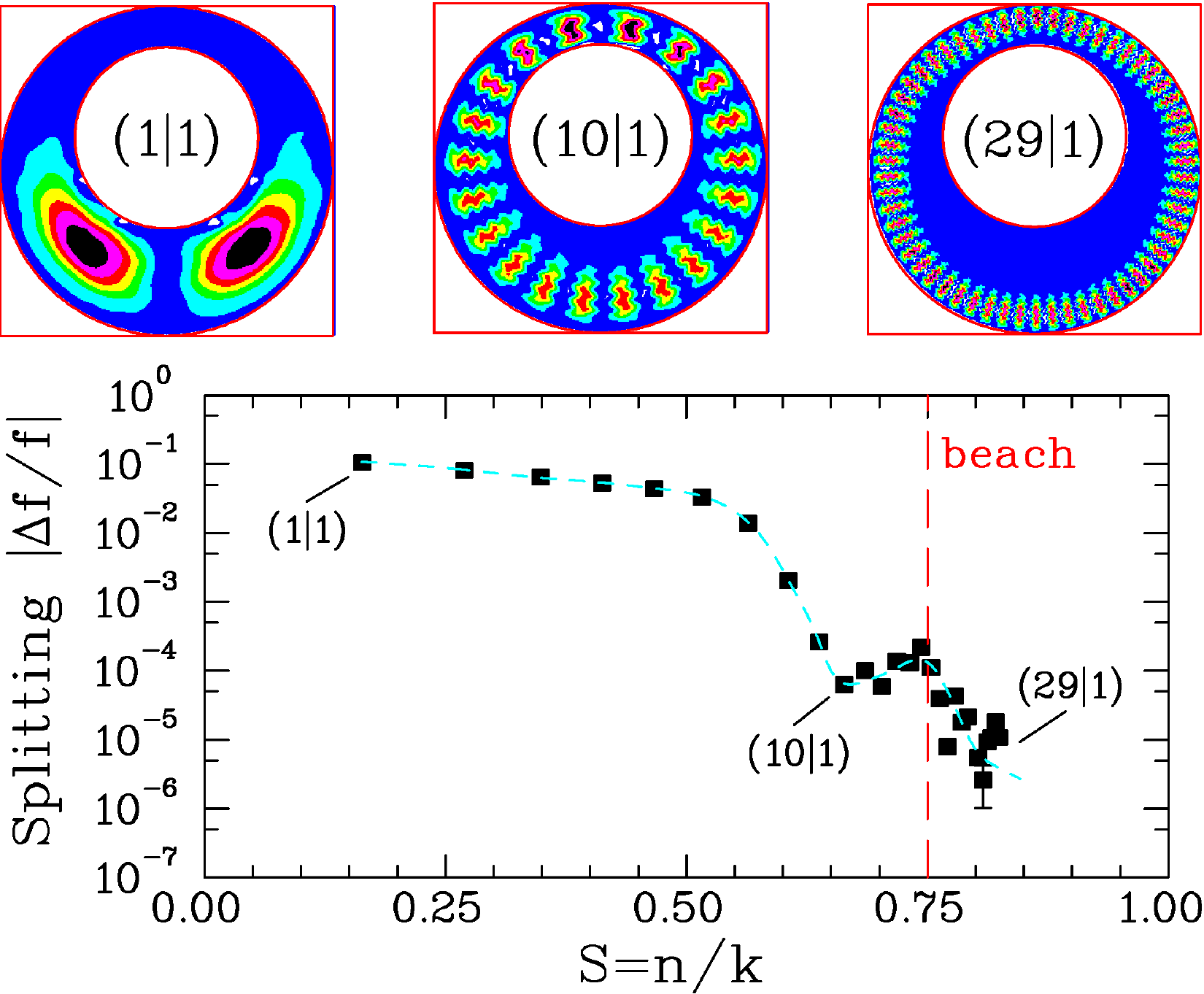}
\caption{(Color online) Splittings of the doublets of resonances (squares) with respect to their average resonance frequency versus its position in the PSOS. The error in the determination of the frequencies of the doublet partners is generally smaller than the size of the squares except for the smallest splitting observed. A smooth transition from large splittings in the chaotic sea ($S<0.75$) towards very small ones in the regular coastal region ($S>0.75$) is visible. The appearance of a pronounced maximum in the direct vicinity of the beach ($S\simeq 0.75$) is a clear indicator for the occurrence of chaos-assisted tunneling.  Reprinted from Phys. Rev. Lett. {\bf 84}, 867 (2000).}
\label{splitbeachannular}
\end{figure}

We would like to close this section by mentioning another experiment on dynamical tunneling, which we performed $~15$~years ago with superconducting microwave billiards with the shapes of annular billiards~\cite{Dembowski2000a,Hofferbert2005}. There, the dynamical tunneling between quasimodes corresponding to clockwise and anticlockwise moving whispering gallery orbits, respectively, was investigated. As recognizable in the upper panels of Figs.~\ref{splitannular} and~\ref{splitbeachannular}, such a billiard corresponds to a circular one of radius $R$ that contains a circular disk of radius $r<R$, where the circle centers are displaced by a value $\delta$ with respect to each other. The classical dynamics is characterized by the value $(r+\delta)/R$ which was chosen equal to $0.75$. Then the regions of neutrally stable clockwise and anticlockwise moving whispering gallery orbits are located in the PSOS at angular momentum values $S>0.75$ and $S<-0.75$ in units of $R$, respectively, where $S=\sin\theta$ corresponds to $p$ in Fig.~\ref{psscw}. For $\vert S\vert=\vert\sin\theta\vert <0.75$ the PSOS consists of a large chaotic sea with a regular island at its center. 

The objective of the experiments was the investigation of chaos-assisted tunneling~\cite{Bohigas1993,Tomsovic1994} between quasimodes localized in the regions $\vert S\vert > 0.75$ via the chaotic sea, that is, the enhancement of the tunneling through a coupling of these modes to chaotic ones. The coupling is largest for modes located close to the borders $S\simeq\pm 0.75$ that separate the three regions. As in the case of the constant-width billiard the tunneling manifests itself in the splitting of the degenerate quasimodes corresponding to clockwise and anticlockwise moving whispering gallery orbits into doublets. Thus, if chaos-assisted tunneling takes place, its size should depend on their location in the PSOS, i.e., on their angular momentum values. According to the Einstein-Brillouin-Keller quantization the quantum analogue of the angular momentum $S=\sin\theta$ is given as $S=n/k_{n,m}$ for a pair of whispering gallery modes. Here, $k_{n,m}$ denotes their mean wave number and their angular and radial quantum numbers are $n$ and $m$, respectively. They actually coincide to a good approximation with those of the integrable annular billiard with $\delta =0$. To determine them the electric field distributions, i.e. the wave functions of the annular billiard, were measured in a normal conducting microwave billiard made from copper using the perturbation method~\cite{Dembowski2000a,Maier1952}. The mean wave numbers $k_{n,m}$ of the whispering gallery modes and the associated splittings were obtained from resonance spectra that were taken with superconducting billiards constructed from niobium. 
Figure~\ref{splitannular} shows a small part of a transmission spectrum which contains two singlets and one doublet, which is assigned to a pair of whispering gallery modes with the angular quantum number $n=18$ given by half the number of maxima of the field intensity shown in the middle panel above the spectra. Furthermore, it has radial quantum numer $m=1$, since the latter exhibits only one maximum in the radial direction. The singlets correspond to modes that are localized in the region with $\vert S\vert<0.75$. 

For the study of the dependence of the tunneling rates, that is, of the sizes of the splittings on the location of the associated doublet modes in the PSOS, i.e., on their angular momentum values, we considered only modes with radial quantum number $m=1$. Figure~\ref{splitbeachannular} shows the resulting splittings versus the angular momentum $S=n/k_\mu$, with $k_\mu=2\pi f_\mu/c$ denoting the wave number corresponding to the resonance frequency $f_\mu$ of the mode. The curve exhibits a clear peak in the splittings, i.e., an enhancement of the tunneling in the beach region around the border $\vert S\vert=0.75$ between the regular coast and the chaotic sea. To the best of our knowledge, this corroborated for the first time the occurrence of chaos-assisted tunneling.  
\section{\label{Dirac} Spectral properties of superconducting microwave Dirac billiards modelling graphene}
Only recently we have started a series of experiments with microwave photonic crystals~\cite{Yablonovitch1989} that model properties of graphene~\cite{Beenakker2008,Castro2009}, a monoatomic layer of carbon atoms arranged on a honeycomb lattice. Due to its particular electronic properties, graphene attracted a lot of attention over the last years. Its conductance and its valence band form conically shaped valleys that touch each other at the six corners of the hexagonal Brillouin zone. In their vicinity the electron energy depends linearly on the quasimomentum, implying an energy independent velocity. Thus, the electrons and holes behave like massless relativistic particles governed by the Dirac equation for spin-$1/2$ particles. Accordingly, the touching points are referred to as Dirac points~\cite{Beenakker2008,Castro2009}. 

This extraordinary feature of the band structure of graphene results in a number of electronic properties which have an analogue in relativistic quantum mechanics. It actually stems from the symmetry properties of its honeycomb structure which is formed by two interpenetrating triangular lattices with threefold symmetry. This led to the manufacturing of artificial graphene using two-dimensional electron gases, molecular assemblies, ultracold atoms~\cite{Singha2011,Nadvornik2012,Gomes2012,Tarruell2012,Uehlinger2013} or photonic crystals~\cite{Bittner2010,Kuhl2010,Sadurni2010,Bittner2012,Bellec2013,Rechtsman2013,Rechtsman2013a,Khanikaev2013}. We exploited this fact to simulate the spectral properties of graphene with microwave resonators that were constructed by squeezing a photonic crystal, which is composed of several hundreds of metallic cylinders arranged on a triangular lattice, between two metal plates. The shape of its band structure, i.e., of its frequencies of wave propagation as function of the two components of the quasimomentum vector, is equivalent to that of the conductance and the valence band in graphene, e.g., for its 1st and its 2nd band. Accordingly, around the Dirac points the propagation of the microwaves in the photonic crystal can be effectively described by the Dirac equation for massless spin-$1/2$ particles. Such a region is identified in the resonance spectrum as one of an exceptionally low resonance density, which increases linearly with the distance from the frequency of the Dirac point, the so-called Dirac frequency~\cite{Bittner2010,Kuhl2010}. 

\begin{figure}[h!]
\includegraphics[width=0.4\linewidth,height=0.25\linewidth]{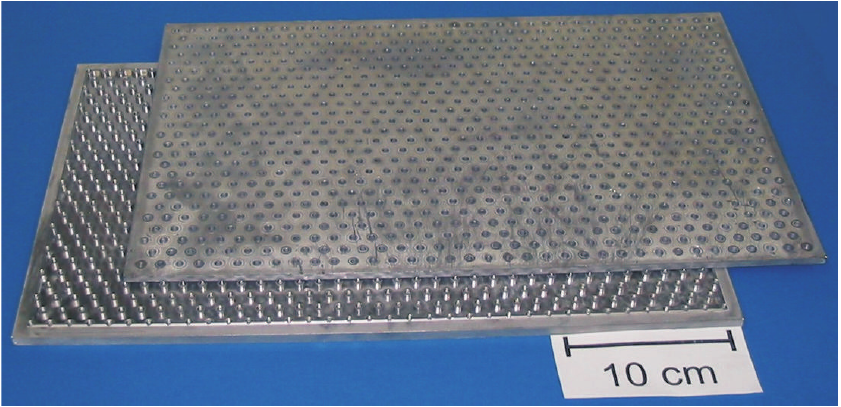}
\includegraphics[width=0.4\linewidth,height=0.25\linewidth]{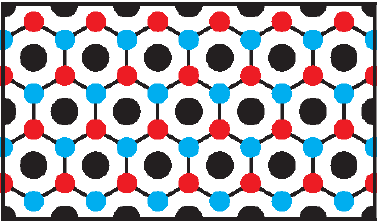}
\caption{(Color online) Left Part: Photograph of the Dirac billiard with the shape of a rectangle, which contains $888$ metal cylinders arranged on a triangular grid. The top plate is shifted with respect to the bottom one.  Reprinted from Phys. Rev. B {\bf 88}, 104101 (2013). Right Part: Schematic view of the triangular lattice of the rectangular Dirac billiard. The red (gray) and blue (dark gray) dots mark the voids between the cylinders. They form a honeycomb lattice.}
\label{photosdiracrect}
\end{figure}

\begin{figure}[h!]
\includegraphics[width=0.7\linewidth]{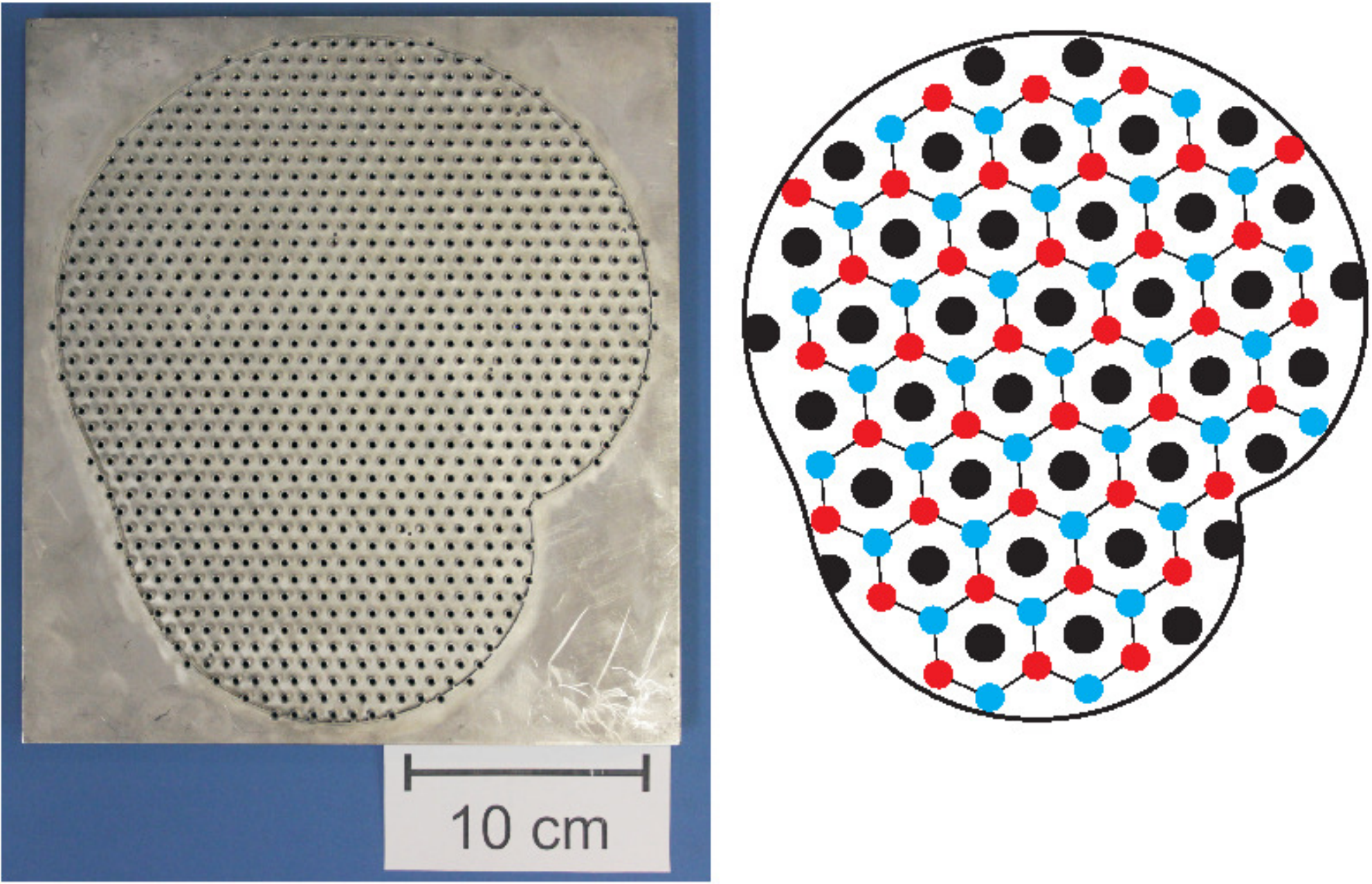}
\caption{(Color online) Same as in Fig.~\ref{photosdiracrect} but for an Africa billiard. The top plate is removed.}
\label{photosdiracafr}
\end{figure}

In order to investigate this behavior in detail, we performed high-precision experiments with superconducting Dirac billiards with the shapes of a rectangle and the African continent~\cite{Berry1987,Huang2010,Huang2011}, respectively. Photographs are shown in the left parts of Figs.~\ref{photosdiracrect} and~\ref{photosdiracafr}. We use them to simulate properties of graphene billiards or graphene flakes. Both billiards consist of a bassin with the shape of the billiard and a lid, that were made from brass plates and then lead plated to attain superconductivity at liquid helium temperature. For the construction of the photonic crystals located inside the bassins $\approx 900$ cylinders were milled out of a brass plate. The height of the resonators was $d=3$~mm so for $f\leq f_{max}=50$~GHz the associated Helmholtz equation is mathematically identical to the Schr\"odinger equation of the quantum billiard containing circular scatterers at the positions of the cylinders. 

The great advantage of Dirac billiards as compared to graphene flakes is that we can control the boundary conditions. Here, those for the wave functions at the voids that are located at the edges of the photonic crystal are of relevance. They can be manipulated by varying the relative position between these voids and the walls of the microwave billiard, where the electric field strength vanishes, i.e., obeys the Dirichlet boundary condition. 

The cylinders were arranged on a triangular grid, as indicated in the schematic views in Figs.~\ref{photosdiracrect} and~\ref{photosdiracafr}, so the voids at the centers of the cells formed by, respectively, three of them yield a hexagonal configuration similar to the honeycomb structure in graphene. Each of the cells may host a quasibound state and thus can be considered as one of the atoms forming the two independent triangular sublattices in graphene. Accordingly, both graphene and microwave photonic crystals can be treated in a similar way within the tight-binding model (TBM). 

\begin{figure}[h!]
\includegraphics[width=0.7\linewidth]{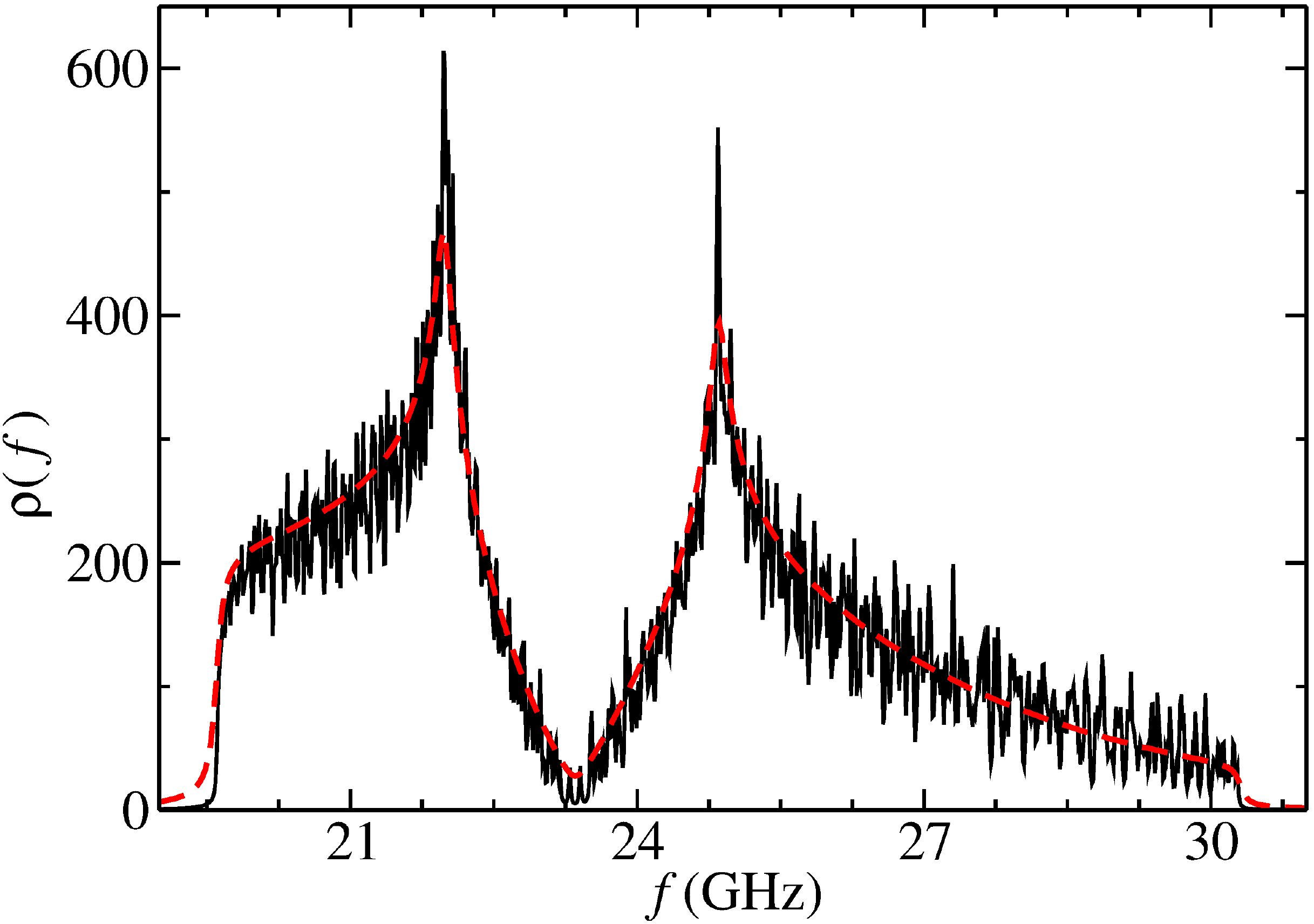}
\includegraphics[width=0.7\linewidth]{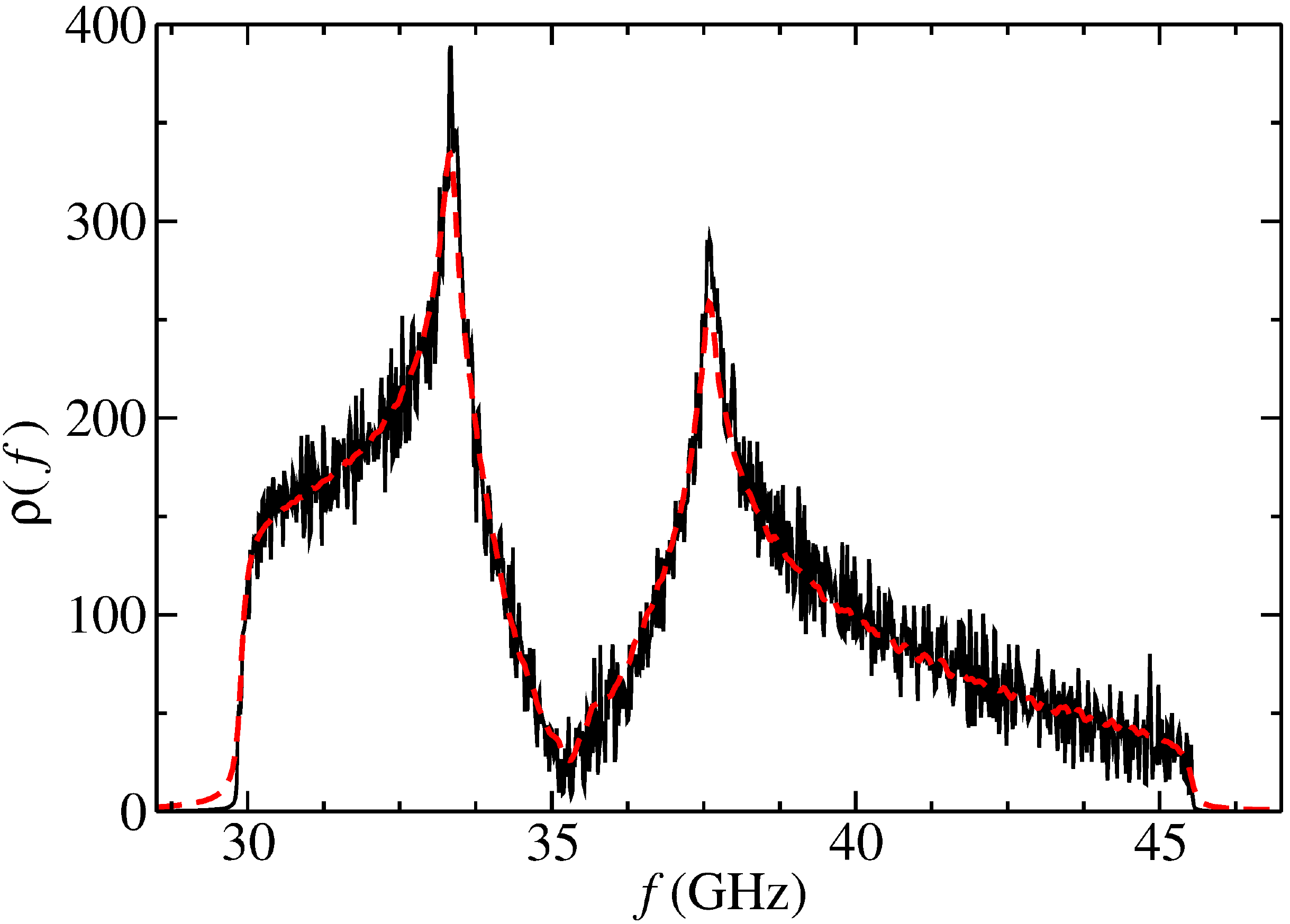}
\caption{(Color online) Densities of the resonance frequencies of the Dirac billiards with the shape of a rectangle [upper panel] and of the African continent [lower panel]. A slight bump is visible in the latter to the right of the Dirac frequency. It is attributed to edge states occurring along the zigzag edge parts of the void structure, see also the right hand side of Fig.~\ref{photosdiracafr}.}
\label{rhosdirac}
\end{figure}

We determined altogether $1651$ and $1823$ resonance frequencies in the first two bands of the rectangular and the Africa Dirac billiard, respectively. Figure~\ref{rhosdirac} shows their resonance densities (black curves). They clearly differ from those expected for the corresponding \emph{empty} microwave billiards, that, according to Weyl's law, would increase linearly with the resonance frequency. Both exhibit a minimum at the frequency of the Dirac point, which separates the two bands. It is bordered by two sharp peaks corresponding to van Hove singularities, that generally occur in two-dimensional crystals with a periodic structure~\cite{Van1953}. There, the resonance density diverges logarithmically for infintely extended crystals, whereas in the case of the Dirac billiards the peak heights increase logarithmically with the size of the crystal, i.e. with the number of voids formed by the cylinders~\cite{Dietz2013}. 

The red (gray) dashed lines in both panels of Fig.~\ref{rhosdirac} correspond to fits of the TBM~\cite{Reich2002} for finite-size photonic crystals to the experimental ones. It incorporates nearest-neighbor, as well as second- and third-nearest-neighbor couplings and in addition the corresponding overlaps of the wave functions centered at the different voids. 
Furthermore, the resonance density of the Africa Dirac billiard exhibits a slight bump to the right of the Dirac frequency, which is attributed to the zigzag edge parts of the void structure formed by the associated photonic crystal~\cite{Wurm2011,Dietz2015,Dietz2015a}. The voids of the rectangular billiard also have a zigzag structure along its longer sides, see Fig.~\ref{photosdiracrect}, however, the photonic crystal can be reproduced periodically via reflections at all its sides so no edge states are generated~\cite{Dietz2015}.      

We demonstrated recently in Ref.~\cite{Dietz2015} that the van Hove singularities divide the two bands framing the Dirac point into two frequency ranges below and above them, where the system is described by the nonrelativistic Schr{\"o}dinger equation of the \emph{empty} quantum billiard, and one between them, where the underlying wave equation coincides with the Dirac equation of the graphene billiard or flake of corresponding shape. Actually, the TBM does not only provide the eigenvalues, but also the wave functions of a bounded photonic crystal. In the regions close to the lower and the upper band edges, they are \emph{identical} with those of the corresponding quantum billiard. Moreover, we showed that this also holds for the fluctuating part of the integrated resonance density and for the spectral fluctuation properties. Furthermore, the computation of the momentum distributions using the wavefunctions and also the direct comparison of the band structure of the photonic crystals with the resonance frequencies plotted versus the eigenvalues of the quantum billiard yielded, that in the vicinity of the lower and the upper band edge the latter maybe identified with the quasimomenta. Accordingly, we call the regions below and above the van Hove singularities the nonrelativistic Schr{\"o}dinger region. 

In the region between the van Hove singularities, i.e., around the Dirac point, the electromagnetic waves inside the billiards are governed by the Dirac equation. Accordingly, there we compared the resonance frequencies and their fluctuation properties with those of the eigenvalues of the corresponding graphene billiard~\cite{Wurm2011}. The latter were computed by solving the Dirac equation with appropriate boundary conditions depending on the edge structure. We arrived at the conclusion, that they and the resonance frequencies interdepend linearly and furthermore, that they coincide with the quasimomenta in the vicinity of the Dirac point.      
Thus, at the van Hove singularities a transition takes place between a nonrelativistic Schr\"odinger region and the relativistic Dirac region, where the Dirac billiard exhibits the same features as the corresponding \emph{empty} quantum billiard and the graphene billiard, respectively. In Ref.~\cite{Dietz2013} we provided evidence that it can be interpreted as a Lifshitz and an excited state quantum phase transition. Presently, we perform more investigations concerning the spectral properties of the Dirac billiards at the van Hove singularities in order to corroborate this assumption. 

\begin{figure}[h!]
\includegraphics[width=0.6\linewidth]{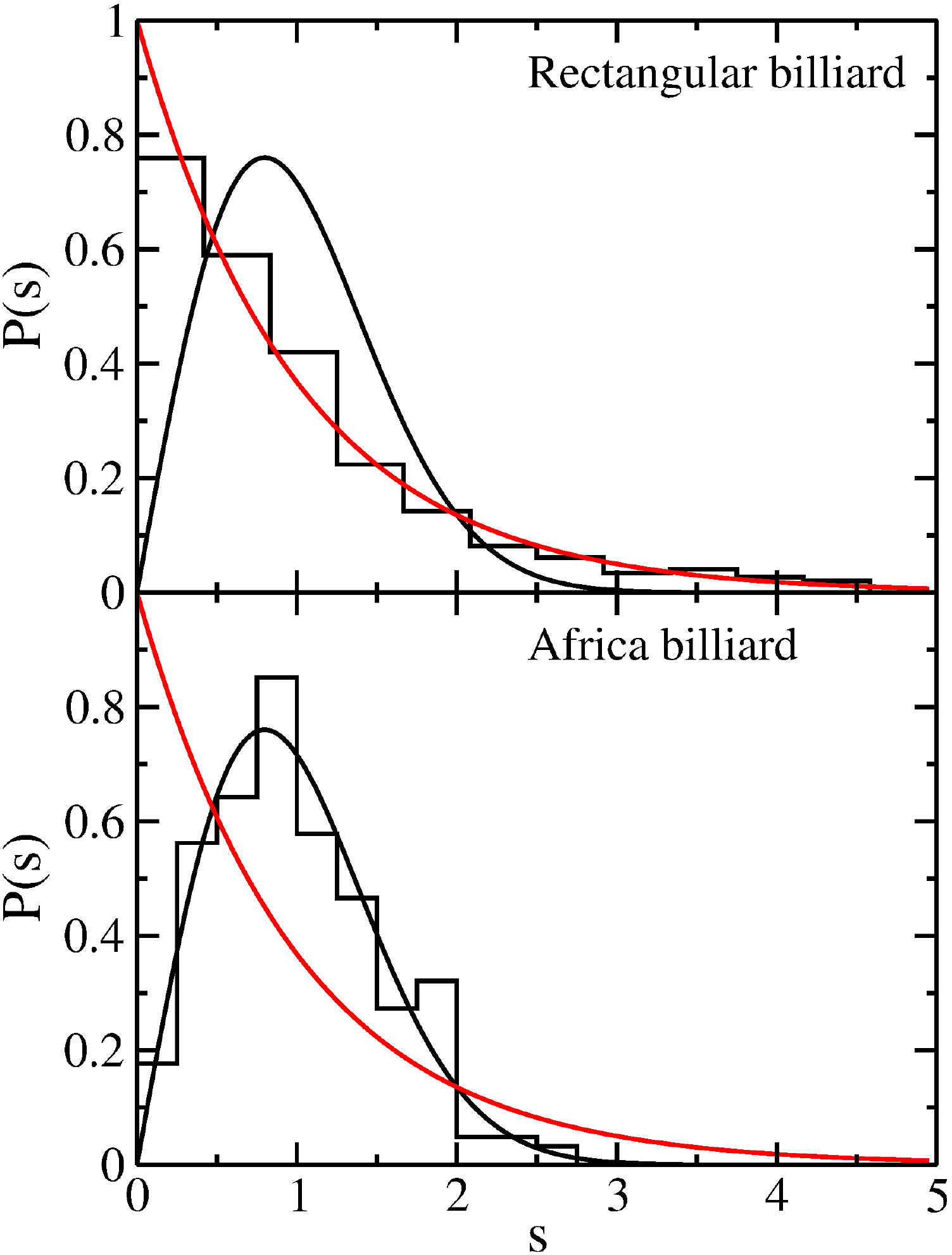}
\caption{(Color online) Nearest-neighbor spacing distributions of the resonance frequencies obtained for the Dirac billiards (histograms) with the shapes of a rectangle (upper panel) and of the African continent (lower panel). The former agrees well with the Poisson distribution (red [gray] line), the latter with the RMT result for the GOE (black line). The upper figure has been reprinted from Phys. Rev. B {\bf 91}, 035411 (2015).}
\label{abstdirac}
\end{figure}

\begin{figure}[h!]
\includegraphics[width=0.6\linewidth]{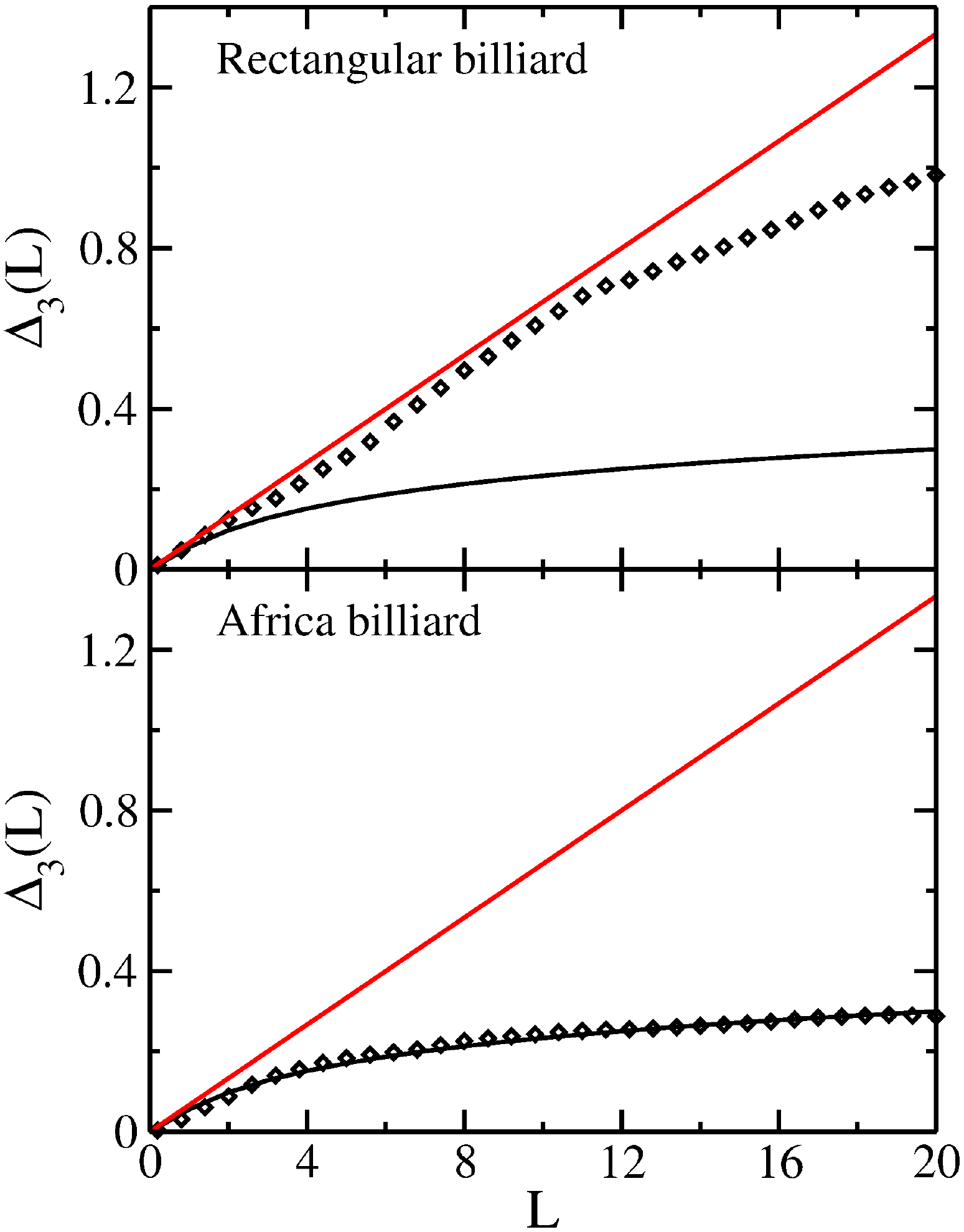}
\caption{(Color online) The $\Delta_3$ statistics of the experimentally determined resonance frequencies of the Dirac billiards (diamonds) with the shapes of a rectangle (upper panel) and of the African continent (lower panel). The former agrees well with that of Poissononian random numbers, the latter with the RMT result for the GOE. The upper figure has been reprinted Phys. Rev. B {\bf 91}, 035411 (2015).}
\label{delta3dirac}
\end{figure}

The spectral properties are similar in the different regions and agree with those of the corresponding \emph{empty} quantum billiard. The classical dynamics of the \emph{empty} rectangular and Africa billiards is regular, respectively, chaotic~\cite{Berry1987,Huang2010,Huang2011}. The nearest-neighbor spacing distribution and the $\Delta_3$ statistics of the rectangular Dirac billiard shown in the upper panels of Fig.~\ref{abstdirac} (histogram) and Fig.~\ref{delta3dirac} (diamonds), respectively, are well described by those of Poissonian random numbers (red [gray] lines) whereas those of the Africa one plotted in the lower panels coincide with the GOE results (black line).

\section{\label{Concl}Conclusions}
The objective of this article was to present some of the highlights of our experimental investigations concerning the quantum manifestations of classical chaos. The focus has been set on spectral fluctuation properties, on symmetry breaking showing up in particular properties of the eigenvalues and the wave functions, on the phenomenon of dynamical tunneling, and, moreover, the universal properties of graphene. Due to the limited space, we refered to only a few examples of the numerous experiments performed with superconducting microwave cavities in the last two decades. Accordingly, we selected some that provided groundbreaking new insight into physical phenomena. Future projects concern the realization of scattering systems by using microwave billiards with one or several openings in the side walls~\cite{Doron1990,Dettmann2009,Altmann2013}. The objective is the measurement of the temporal decay of microwave power through these openings and a comparison with RMT and other predictions. Such experiments yield information on the universal decay behavior of open systems like excited nuclei. In order to ensure that the system decays exclusively via the openings, the dissipation into the walls of the resonator needs to be eliminated. This is only possible by performing the measurements of the $S$-matrix elements at superconducting conditions. There, however, a calibrated measurement, an essential requirement for the analysis of properties of the $S$ matrix, is not yet feasible. Such a method was developed only recently~\cite{Yeh2013}. Currently we implement and adjust it to our superconducting measurements of the $S$ matrix with open microwave resonators with the shapes of regular and chaotic billiards. The aim is to obtain information on the differences in their decay behavior. Furthermore, we construct superconducting quantum graphs~\cite{Kottos1997}, a network-shaped structure of vertices bonded by rectangular waveguides. The objective of these experiments is the experimental test of RMT predictions concerning the correlation functions of the $S$ matrix of open quantum graphs~\cite{Pluhar2013,Pluhar2014} in high-precision experiments. Finally, a further current project concerns the experimental verification of the so-called Atiyah-Singer index theorem~\cite{Atiyah1968,Atiyah1968a,Atiyah1968b,Pachos2007}, which relates the number of zero modes of the Dirac operator to the topology of the associated graphene sheet. The resonator consists of a network of rectangular waveguides that connect circular cavities. It was milled out of a brass sphere and has the form of a fullerene C$_{60}$ molecule~\cite{Kroto1985}. Summarizing, we have with our macroscopic, superconducting microwave resonators systems available that might be tailored to model the universal properties of a large range of microscopic quantum systems in high precision experiments. 
\begin{acknowledgments}
This work was supported by the Deutsche Forschungsgemeinschaft (DFG) within the Collaborative Research Center 634. We thank all the co-authors of our publications quoted in the list of references for a fruitful collaboration and for their many contributions to the scientific output of the experiments presented in this article. From the beginning the late Oriol Bohigas and Martin Gutzwiller have accompanied our work on superconducting billiards with great interest and support. We thus gratefully dedicate this review to their memory.     
\end{acknowledgments}

\end{document}